\documentclass{article}

\usepackage{arxiv}

\usepackage[utf8]{inputenc} % allow utf-8 input
\usepackage[T1]{fontenc}    % use 8-bit T1 fonts
\usepackage{hyperref}       % hyperlinks
\usepackage{url}            % simple URL typesetting
\usepackage{booktabs}       % professional-quality tables
\usepackage{amsfonts}       % blackboard math symbols
\usepackage{nicefrac}       % compact symbols for 1/2, etc.
\usepackage{microtype}      % microtypography
\usepackage{lipsum}		% Can be removed after putting your text content
\usepackage{graphicx}
\usepackage[super,sort&compress]{natbib}
\usepackage{doi}
\usepackage{amsmath,amssymb,mathtools}
\title{Robust bootstrap methods for meta-analyses of diagnostic test accuracy studies}

\author{ 
	\href{https://orcid.org/0000-0002-0171-588X}{\includegraphics[scale=0.06]{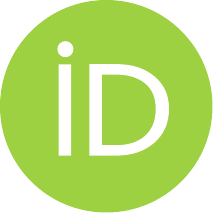}\hspace{1mm}Diribsa T. Bedada} \\
	Department of Statistics and Actuarial Science\\
	University of Waterloo\\
	Waterloo, Ontario, Canada \\
\And
Olana A. Dabi \\
	Department of Statistics and Actuarial Science\\
	University of Waterloo\\
	Waterloo, Ontario, Canada \\
\And
Isaac Fordjour \\
	Department of Statistics and Actuarial Science\\
	University of Waterloo\\
	Waterloo, Ontario, Canada \\
\And
\href{https://orcid.org/0000-0002-3154-3855}{\includegraphics[scale=0.06]{orcid.pdf}\hspace{1mm}Zelalem F. Negeri}\thanks{Corresponding author: Zelalem F. Negeri, Ph.D., Department of Statistics and Actuarial Science; University of Waterloo, M3 4116; Waterloo, Ontario, Canada N2L 3G1; Tel +1 519 888 4567, ext 4118; Email: znegeri@uwaterloo.ca.} \\
	Department of Statistics and Actuarial Science\\
	University of Waterloo\\
	Waterloo, Ontario, Canada
}

\renewcommand{\shorttitle}{Bootstrap methods for meta-analysis of DTAs}

\begin{document}
\maketitle

\begin{abstract}
Meta-analysis of diagnostic test accuracy (DTA) commonly relies on the bivariate linear mixed model (BLMM) and the bivariate generalized linear mixed model (BGLMM) to jointly synthesize diagnostic test sensitivity ($Se$) and specificity ($Sp$). However, reliance on asymptotic theory and random-effects distributional assumptions may affect finite-sample inference, particularly in small meta-analyses or when the random effects are non-normal. We proposed three bootstrap-based approaches: a nonparametric bootstrap (NPB), a parametric bootstrap (PB); and a bias-corrected parametric bootstrap (BCPB), with the latter two based on a fitted BGLMM. NPB model parameters were estimated using inverse-variance-covariance weighted least squares. Extensive simulations under bivariate normal, Student-t, and Laplace random-effects distributions evaluated finite-sample performance and robustness to departures from normality. Results were qualitatively similar across the three distributions. The bootstrap approaches generally outperformed BLMM in terms of bias, root mean squared error (RMSE), and coverage probability, particularly when diagnostic accuracy is high and the number of studies is small. The BCPB demonstrated the strongest finite-sample performance among the bootstrap approaches, frequently achieving the lowest bias and RMSE for pooled sensitivity and specificity while maintaining coverage probabilities close to the nominal level. The BGLMM also performed well across most settings. The NPB approach generally improved upon the BLMM while providing a flexible alternative without parametric random-effects assumptions. Real-data applications further illustrated the practical utility of the proposed approaches. Overall, the bootstrap methods provide useful alternatives for DTA meta-analysis, particularly when the number of studies is small or random-effects distributional assumptions are uncertain.
\end{abstract}

% keywords can be removed

\section{Introduction}\label{s_1}
Accurate and timely diagnostic tests are crucial for effective healthcare, as they enable early detection, appropriate treatment planning, monitoring of disease progression, and assessment of treatment effectiveness. A diagnostic test is a technique that identifies individuals with respect to their target condition, for example, disease status (diseased or non-diseased). \cite{negeri_bivariate_2018}  Meta-analysis is a statistical method used in research to analyze and combine evidence from multiple independent studies addressing the same research question or topic.\cite{borenstein2009effect,glass_primary_1976} It has been shown that hierarchical linear mixed models are useful in meta-analyzing DTA studies.\cite{raudenbush_empirical_1985,noortgate_parametric_2005}  Meta-analyzing DTA studies is a type of meta-analysis in which the DTA characteristics, such as $Se$ and $Sp$, are synthesized by combining and analyzing data from multiple independent studies to evaluate the performance of diagnostic tests. \cite{negeri2019contributions, negeri_robust_2020, negeri_statistical_2020}

Over the past two decades, several parametric bivariate hierarchical models that account for potential associations between $Se$ and $Sp$ and between-study variation have been proposed to assess the accuracy of diagnostic tests. Each of these models has its own strengths and limitations. Among the models used to synthesize DTA measures, primarily the logit-transformed $Se$ and $Sp$, the approaches proposed by Reitsma  \textit{et al.} (2005), \cite{reitsma2005bivariate} and Chu \& Cole (2006) \cite{chu_bivariate_2006} have been the most applied in the current literature. 

Reitsma  \textit{et al.} (2005) \cite{reitsma2005bivariate} proposed a bivariate random-effects model, which we refer to as the BLMM, that jointly models the logit-transformed $Se$ and $Sp$ by assuming that the random effects follow a bivariate normal distribution. In this method, the study-specific variability in $\text{logit}(Se)$ and $\text{logit}(Sp)$ is modelled by an approximate (asymptotic) bivariate normal distribution. The BLMM is implemented in the \texttt{mada} package of the R programming language. \cite{doebler_mada_2025} Chu and Cole (2006) \cite{chu_bivariate_2006} developed a generalized bivariate random-effects model we refer to as the BGLMM that avoids the need to add an ad hoc Haldane-Anscombe correction \cite{haldane1956estimation,anscombe1956estimating} to DTA data containing zero cell counts. The BGLMM meta-analyzes observed binary diagnostic test outcomes from each study using a generalized linear mixed model. The key advantage of the BGLMM by Chu and Cole (2006) over the BLMM of Reitsma  \textit{et al.} (2005) is that, instead of modelling the observed within-study variability by an approximate normal distribution, the BGLMM models it exactly using the Binomial distribution. Despite its attractive statistical properties, the BGLMM relies on likelihood-based estimation procedures that may lead to convergence issues in small meta-analyses or sparse-data settings. Negeri et al. (2018)\cite{negeri_bivariate_2018} proposed the arcsine square root and Freeman--Tukey double arcsine transformations for bivariate diagnostic meta-analysis of DTA studies, eliminating the need for the Haldane--Anscombe correction. However, these methods still rely on asymptotic within-study variance estimators, which are appropriate only for large within-study sample sizes. To address this limitation, Dabi and Negeri\cite{dabi2026diagnostic} proposed an exact (analytical) within-study variance calculation approach that does not require a Haldane–Anscombe correction and remains invariant to transformations. However, their approach, which uses a logit transformation, still requires a Haldane–Anscombe correction for the smallest and largest values in the binomial range.

The parameters of these standard models are often estimated using maximum likelihood, and inferences are based on asymptotic theory. Unfortunately, this may yield biased parameter estimates and standard errors, as maximum likelihood estimation in hierarchical linear models is problematic.\cite{VanDenNoortgate2005} The estimated standard errors and the normal sampling distribution assumptions for the maximum likelihood estimates apply only asymptotically. Therefore, these assumptions may be violated in small samples. Although the sampling distributions of the fixed-effects coefficients are often well approximated by $t$ distributions, the sampling distributions of the variance components are typically skewed to an unknown degree, and their estimates may be biased in small samples. \cite{bryk_hierarchical_nodate,negeri_skew-normal_2020, noortgate_parametric_2005} Moreover, the use of maximum likelihood theory can yield misleading results when the normality assumption at each level is violated. \cite{bryk_hierarchical_nodate,negeri_skew-normal_2020} The bivariate normal-skew-normal (BNSN) model was proposed by Negeri and Beyene (2020) \cite{negeri_skew-normal_2020} as an improved approach for modelling the logit-transformed $Se$ and $Sp$ when the distributional assumptions are violated. Through simulation studies, the authors demonstrated that the BNSN model can provide improved estimation and inference when the assumption of normally distributed random effects is violated. However, this and other flexible parametric methods do not adequately account for small numbers of studies or small within-study sample sizes.

To overcome some of these limitations, copula-based bivariate beta-binomial models have been proposed by Kuss et al.(2014).\cite{kuss2014meta} These models directly specify marginal beta-binomial distributions for the numbers of true positives and true negatives and employ copula functions to model their dependence. By avoiding the normality assumption for the random effects, copula models provide greater flexibility in modelling the association between \(Se\) and \(Sp\) and admit a closed-form likelihood. Nevertheless, fitting copula-based models remains computationally challenging, and convergence problems have been reported in both simulation studies and real-world applications. Motivated by these challenges, Zapf  \textit{et al.} (2015) \cite{zapf_nonparametric_2015} proposed an NPB approach for the meta-analysis of DTA studies. Unlike existing parametric methods, the proposed NPB approach do not require distributional assumptions about latent random effects or a prespecified dependence structure between \(Se\) and \(Sp\). 
Consequently, it provides a flexible and computationally robust alternative that remains applicable even in settings where parametric models struggle to converge, such as meta-analyses with few studies or sparse data. However, the nonparametric model proposed by Zapf  \textit{et al.} (2015)\cite{zapf_nonparametric_2015} estimated the primary model parameters ($Se$ and $Sp$) using simple averaging. The challenge with simple averaging is that studies may have different weights, and assigning equal weight to all studies may lead to inaccurate estimates.\cite{negeri2019contributions,negeri_skew-normal_2020}
Bootstrap is a flexible method for obtaining bias-corrected parameter estimates and corresponding standard errors (Efron, 1982).\cite{efron_5_1982} The application of bootstrap methods to hierarchical linear models has previously been discussed in other contexts\cite{goldstein_bootstrapping_2011,laird_empirical_1987} and in meta-analysis. \cite{noortgate_parametric_2005} However, their application to the synthesis of DTA studies has received little attention in the methodological literature. Therefore, this paper aims to address this important research gap by proposing three bootstrap-based approaches: an NPB, which does not require distributional assumptions for the latent random effects; a PB; and a bias-corrected parametric bootstrap (BCPB), which extends the PB approach by incorporating bootstrap bias correction into the point estimation. The proposed bootstrap methods estimate the primary DTA parameters using weighted least squares. Bootstrap methods are expected to be less sensitive to distributional assumptions, provide reliable inference in small meta-analyses, and exhibit improved numerical stability relative to likelihood-based approaches. We evaluate and compare the performance of the proposed methods against existing approaches on both real-world and simulated datasets.

The rest of this article is structured as follows: Section \ref{s_2} presents motivating examples using two real-life datasets. Section \ref{s_3} introduces the statistical methods, including BLMM, BGLMM, NPB, PB, and BCPB. The simulation results and evaluation of the methods using the motivating examples are discussed in Sections \ref{s_5} and \ref{s_6}, respectively. Finally, Section \ref{s_7} provides the discussion and conclusion of the manuscript.

\section{Motivating examples} \label{s_2}
In this Section,  we provide an overview of two real-life datasets to motivate the statistical methods presented in Section \ref{s_3}.

\begin{figure}[ht]
    \centering
    \includegraphics[width=\textwidth]{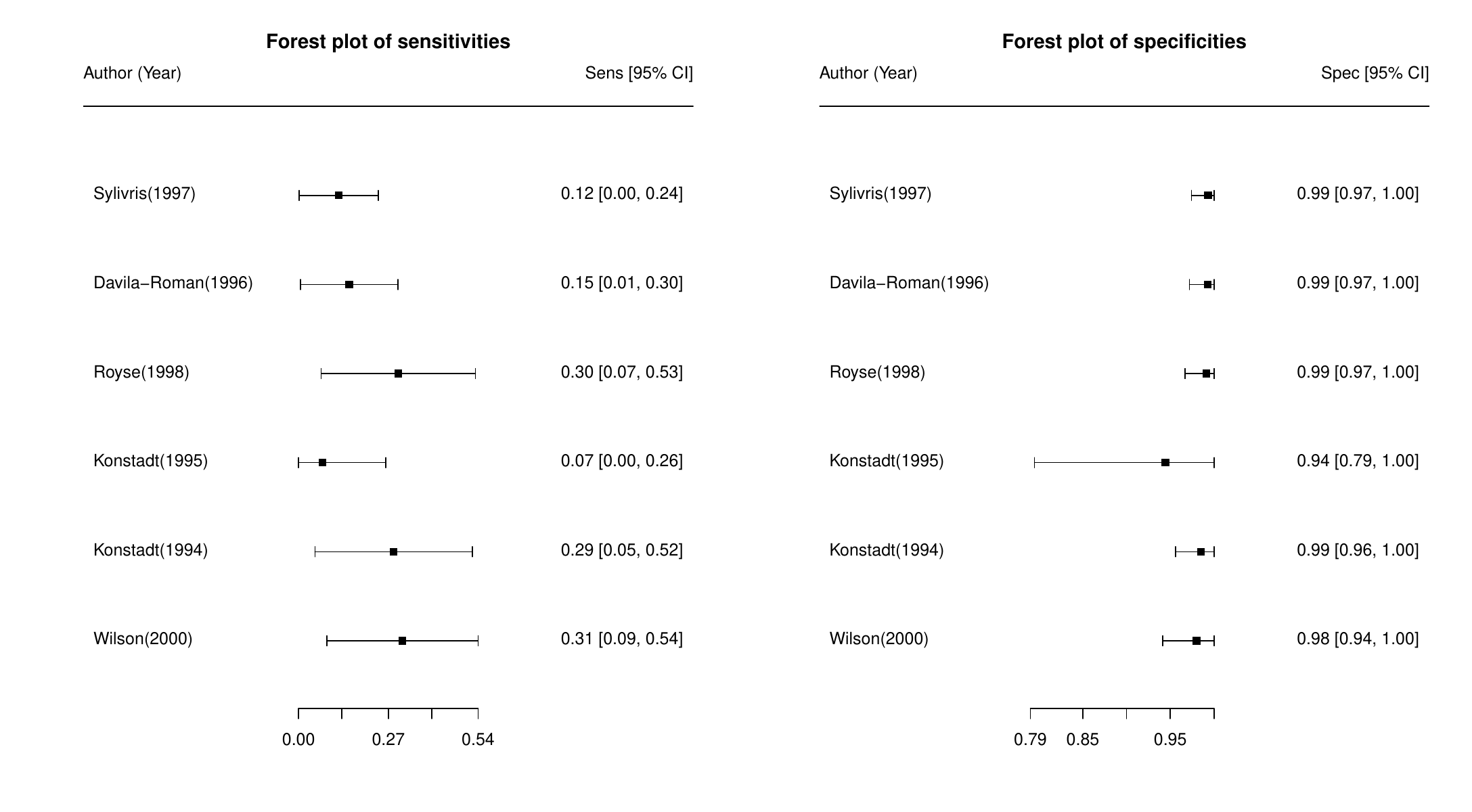} 
    \caption{Forest plot for sensitivity (left) and specificity (right) of the TEE dataset.\cite{van_zaane_meta-analysis_2008}}
    \label{fig_2.1}
\end{figure}

The first real-life dataset, from van Zaane \textit{et al.} (2008)\cite{van_zaane_meta-analysis_2008}, concerns the diagnostic accuracy of transesophageal echocardiography (TEE) for assessing atherosclerosis of the ascending aorta in patients undergoing cardiac surgery. The \texttt{mada}\cite{doebler_mada_2025} R package was used to create the forest plots for the real-life data examples. As shown in the forest plots in Figure \ref{fig_2.1}, this meta-analysis includes only six studies, making it an ideal dataset for evaluating both the existing and proposed methods when it is difficult to justify distributional assumptions. Moreover, the TEE dataset contains, on average, only 17 diseased and 53 non-diseased participants per study.

\begin{figure}[ht]
    \centering
    \includegraphics[width=\textwidth]{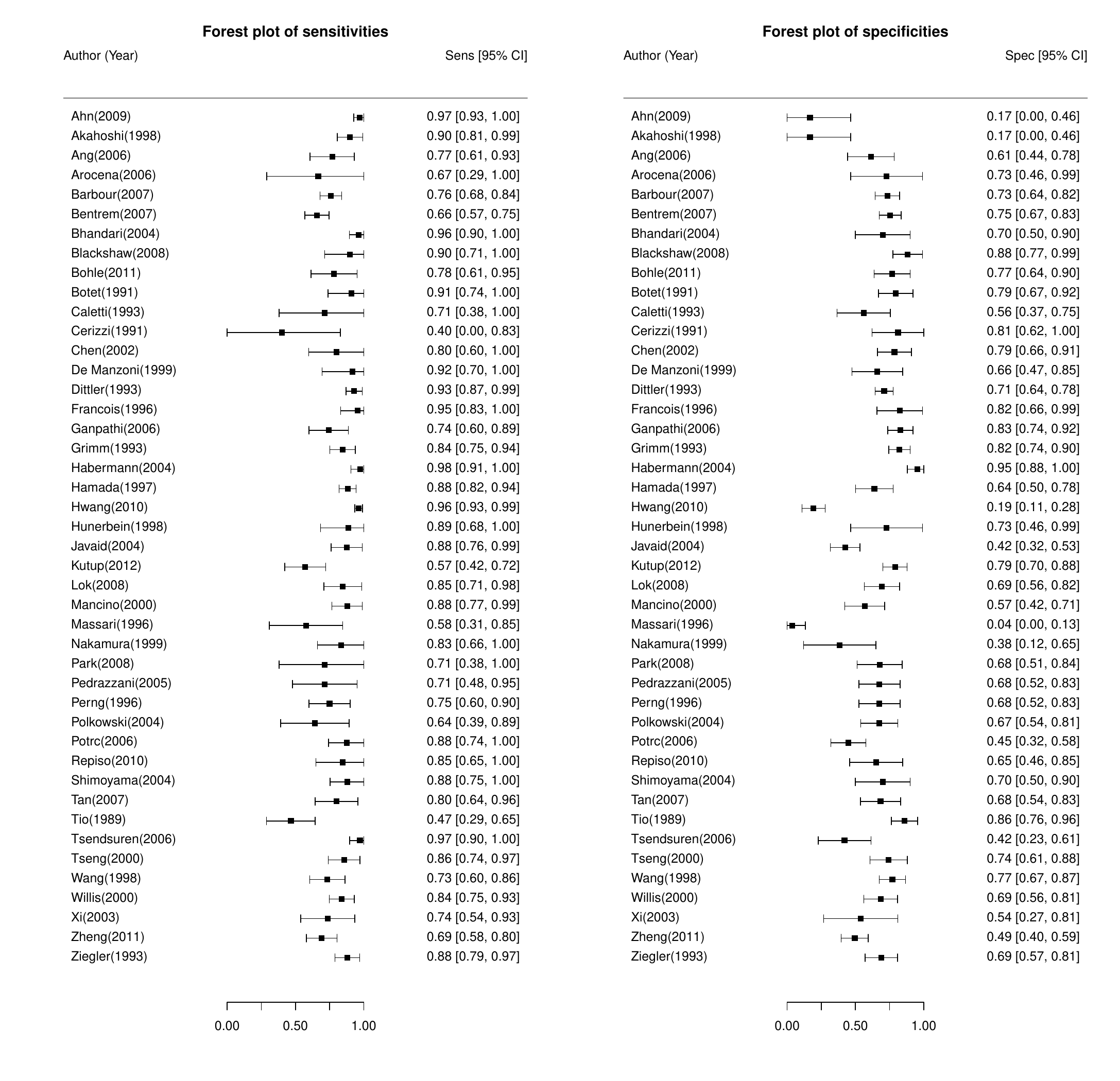} 
    \caption{ Forest plot for sensitivity (left) and specificity (right) of the EUS dataset.\cite{mocellin_diagnostic_nodate}}
    \label{fig_2.2}
\end{figure}

The second real-life dataset is based on the Endoscopic Ultrasonography (EUS) test for the preoperative diagnosis of primary gastric cancer, comprising 44 studies and reported in the Cochrane review by Mocellin and Pasquali (2015).\cite{mocellin_diagnostic_nodate} Negeri and Beyene \cite{negeri_skew-normal_2020} showed that the logit($Sp$) values of the EUS data follow a negatively skewed distribution, whereas the logit($Se$) values are approximately normally distributed. The average numbers of diseased and non-diseased individuals are about 36 and 46, respectively. Figure \ref{fig_2.2} presents the forest plot of the EUS dataset.
We will reanalyze both the EUS and TEE datasets in Section 4 using the standard and proposed approaches.

\section{Methods}
\label{s_3}
In this section, we review two standard bivariate random-effects models and present the proposed bootstrap methods for synthesizing $Se$ and $Sp$ from $k$ independent DTA studies. 

\subsection{Data Structure and Measures of Diagnostic Test Accuracy}

The data structure for diagnostic test studies is presented in a $2 \times 2$ table as summarized in Table \ref{Table_1}. DTA studies commonly report results using two key measures: sensitivity and specificity.\cite{rosenberger2022empirical} Sensitivity, also referred to as the true positive fraction, is the probability that a diagnostic test correctly classifies diseased individuals as positive, whereas specificity, or the true negative fraction, is the probability that the test correctly classifies non-diseased individuals as healthy. In practice, these two measures are often inversely related, meaning that improving sensitivity may reduce specificity, and vice versa.

\begin{table}[ht]
\centering
\caption{ Data Structure of a Diagnostic Test Study}
\label{Table_1}
\begin{tabular}{lcc}
\toprule
& \multicolumn{2}{c}{\textbf{True disease Status}} \\
\cmidrule(lr){2-3}
\textbf{Test Result} & Diseased & Non-diseased \\
\midrule
Positive & True Positive ($TP$) & False Positive ($FP$) \\
Negative & False Negative ($FN$) & True Negative ($TN$) \\
\midrule
Total & $n_1 = TP + FN$ & $n_2 = FP + TN$ \\
\bottomrule
\end{tabular}
\end{table}

Based on the observed data presented in Table \ref{Table_1}, these measures can be estimated by
\begin{equation*}
\widehat{Se}=\dfrac{TP}{n_1}, \qquad
\widehat{Sp}=\dfrac{TN}{n_2},
\end{equation*}
where \(n_1\) and \(n_2\) denote the total number of diseased and non-diseased individuals. Moreover, $TP$ represents the number of true positives, namely diseased individuals correctly classified as positive by the diagnostic test, whereas $TN$ represents the number of true negatives, namely non-diseased individuals correctly classified as negative by the test.

\subsection{The Bivariate Linear Mixed Model}
\label{sec:3.2}
Assuming there are \(k\) studies, indexed by \(i = 1, \dots, k \), the study-specific $Se$  and $Sp$ estimates for a DTA study are given by:
  \begin{equation*}
\widehat{Se}_i=\dfrac{TP_i}{n_{1i}}, \qquad
\widehat{Sp}_i=\dfrac{TN_i}{n_{2i}}.
\end{equation*}

The BLMM model of Reitsma  \textit{et al.} (2005) \cite{reitsma2005bivariate} assumes that $y_{1i}=logit(\widehat{Se}_i)$ and $y_{2i}=logit(\widehat{ Sp}_i)$ follow a bivariate normal distribution and describes their variability hierarchically in two stages.\\

\textbf{The Within-Study Variation:}
In the first stage, the BLMM model assumes a fixed-effect model to describe the within-study variation in $y_{1i}$ and $y_{2i}$ as follows:
\begin{equation}
\label{Eq_1}
\begin{pmatrix}
y_{1i} \\
y_{2i}
\end{pmatrix}
\sim N_2
\left(
\begin{bmatrix}
\mu_{1i} \\
\mu_{2i}
\end{bmatrix},
\boldsymbol{\Psi}_i =
\begin{bmatrix}
\psi_{1i}^2 & 0 \\
0 & \psi_{2i}^2
\end{bmatrix}
\right),
i = 1, \dots , k, 
\end{equation}

where $(\mu_{1i}, \mu_{2i} )^\top$ denotes the study-specific random effects denoting the unknown mean vectors for $(y_{1i}, y_{2i} )^\top$ and $\boldsymbol{\Psi}_i$ is the within-study covariance for $(y_{1i}, y_{2i} )^\top$, which is typically assumed known. In practice, however, $\boldsymbol{\Psi}_i$ is approximated from the observed data using the delta method, as follows: 
\[
\psi_{1i}^2\ \approx \ \dfrac{1}{n_{1i}\cdot \widehat{Se}_i\cdot (1-\widehat{Se}_i) }, \ \ \ \ \ \ \ \ \ \psi_{2i}^2\ \approx \ \dfrac{1}{n_{2i}\cdot \widehat{Sp}_i\cdot (1-\widehat{Sp}_i) }
\]

\textbf{The Between-Study Variation:}
In the second stage, the BLMM assumes the bivariate normal distribution to describe the heterogeneity between $\mu_{1i}$ and $\mu_{2i}$ across studies: 
\begin{equation}
\label{Eq_2}
\begin{pmatrix}
\mu_{1i} \\
\mu_{2i}
\end{pmatrix}
\sim N_2 \left( \boldsymbol{\mu} =
\begin{bmatrix}
\mu_1 \\
\mu_2
\end{bmatrix},
\boldsymbol{\Sigma} =
\begin{bmatrix}
\sigma^2_1 & \sigma_{12} \\
\sigma_{12} & \sigma^2_2
\end{bmatrix}
\right), i = 1, \dots, k,
\end{equation}
where $\boldsymbol{\mu}=(\mu_{1}, \mu_{2} )^\top$ is the overall mean vector of $y_{1i}$ and $y_{2i}$, and $\boldsymbol{\Sigma}$ is the between-study heterogeneity matrix.

Finally, the BLMM is defined by combining equations \ref{Eq_1} and \ref{Eq_2} to obtain the marginal model:

\begin{equation}
\label{Eq_3}
\begin{pmatrix}
y_{1i} \\
y_{2i}
\end{pmatrix}
\sim N_2
\left( \boldsymbol{\mu}=
\begin{bmatrix}
\mu_1 \\
\mu_2
\end{bmatrix}, \boldsymbol{\Sigma_i} = 
\boldsymbol{\Sigma} + \boldsymbol{\Psi}_i =
\begin{bmatrix}
\sigma_1^2 + \psi_{1i}^2 & \sigma_{12} \\
\sigma_{12} & \sigma_2^2 + \psi_{2i}^2
\end{bmatrix}
\right)
\end{equation}

\subsubsection{Parameter Estimation}
The parameters of interest in the BLMM model are estimated by maximizing the following log-likelihood function:
\begin{equation}
 \label{Eq_3}
\ell (\boldsymbol{\mu}, \boldsymbol{\Sigma}) = -\frac{k}{2} \log (2\pi) - \frac{1}{2} \sum_{i=1}^{k} \log |\boldsymbol{\Sigma_i}| - \frac{1}{2} \sum_{i=1}^{k} (\boldsymbol{y}_i - \boldsymbol{\mu})^T \boldsymbol{\Sigma_i}^{-1} (\boldsymbol{y}_i - \boldsymbol{\mu}),
\end{equation}

\noindent where \(\mathbf{y}_i = (y_{1i}, y_{2i})^{\top}\). It follows from Equation \ref{Eq_3} that the maximum likelihood (ML) estimator of \(\boldsymbol{\Sigma}\) does not have a closed-form solution; hence, the ML estimates of \(\boldsymbol{\mu}\) and \(\boldsymbol{\Sigma}\) are obtained iteratively using numerical methods. The BLMM was fitted using the \texttt{mada} package in R \cite{genz2021package}, which implements the Reitsma bivariate random-effects model for the meta-analysis of DTA studies.

Meta-analysis models can be viewed as special cases of the general linear model, characterized by heterogeneous error variances that are assumed to be known and fixed.
\cite{viechtbauer2010conducting} Meta-analytic models employ a two-stage estimation process: initially, the between-study variance-covariance is estimated,\cite{veroniki2016methods}, followed by the calculation of the pooled effect measure via weighted least squares. In this process, the weights are computed as the inverses of the combined between-study and within-study variance-covariance matrices.\cite{viechtbauer2010conducting,borenstein2021introduction}

Consequently, in the two-stage estimation procedure, the first step is to obtain the ML estimate for $\boldsymbol{\Sigma}$ from equation \ref{Eq_3}. Then unbiased weighted least squares estimator of the pooled logit($Se$) and logit($Sp$) is given by:
\begin{equation}
\label{Eq_4}
\widehat{\boldsymbol{\mu}} = \left( \sum_{i=1}^{k} \boldsymbol{\boldsymbol{\widehat{W}_i}} \right)^{-1} 
\left( \sum_{i=1}^{k} \boldsymbol{\widehat{W}_i} \boldsymbol{y}_i \right). 
\end{equation}
where \(\boldsymbol{\widehat{W}_i}\) denotes the inverse of the total variance-covariance matrix, defined as \(\boldsymbol{\widehat{W}_i} = \left(\boldsymbol{\widehat\Sigma} + \boldsymbol{\Psi}_i\right)^{-1} = \boldsymbol{\widehat\Sigma_i}^{-1} \).\\

The variance-covariance matrix used to construct 95\% confidence intervals (CIs) for $\boldsymbol{\mu}$ is given by:
\begin{equation}
\label{Eq_5}
var(\widehat{\boldsymbol{\mu}})  = \left(\sum_{i=1}^{k} \boldsymbol{\widehat{W}_i} \right)^{-1} = \left(\sum_{i=1}^{k}\boldsymbol{\widehat\Sigma_i}^{-1}\right)^{-1} 
\end{equation}
A (\(1-\alpha\))100\% Wald-type CI for \(\boldsymbol{\mu}\) can be obtained by combining equations \ref{Eq_4} and \ref{Eq_5}: \[ \widehat{\boldsymbol{\mu}}\ \pm \ Z_{1-\alpha/2} \cdot \text{se}(\widehat{\boldsymbol{\mu}}),\] where $se$($\widehat{\boldsymbol{\mu}}$) is the standard error of $\widehat{\boldsymbol{\mu}}$, obtained as the square roots of the diagonal elements of the matrix in equation \eqref{Eq_5}\cite{negeri_skew-normal_2020} and $Z_{1-\alpha/2}$ denotes the $(1-\alpha/2)th$ quantile of the standard normal distribution.

\subsection{The Bivariate Generalized Linear Mixed Model }
\label{sec:3.3}
Chu and Cole (2006)\cite{chu_bivariate_2006} proposed the bivariate generalized linear mixed-effects model, which referred to here as the BGLMM. Unlike the BLMM model, which relies on normal approximations, the BGLMM model directly models the variation in the observed counts of true positives and true negatives using independent binomial distributions. The model is specified as follows:
\begin{equation*}
\begin{split}
TP_i \mid \mu_{1i} &\sim \text{Binomial}(n_{1i}, Se_i); \quad y_{1i} = \text{logit}\left(\widehat{Se}_i\right) = \mu_1 + \mu_{1i}, \\
TN_i \mid \mu_{2i} &\sim \text{Binomial}(n_{2i}, Sp_i); \quad y_{2i} = \text{logit}\left(\widehat{Sp}_i\right) = \mu_2 + \mu_{2i}, \\
\boldsymbol{\mu}_i = (\mu_{1i}, \mu_{2i})^\top &\sim N_2(\mathbf{0}, \ \boldsymbol{\Sigma}),
\end{split}
\tag{3.7}
\end{equation*}
Here, \(TP_i\),  \(TN_i\), \(Se_i\),  \(Sp_i\), \(n_{1i}\) and \(n_{2i}\) are as defined earlier.
An advantage of the BGLMM is that it does not require Haldane-Anscombe corrections when cell counts are zero.\cite{rosenberger2022empirical} Furthermore, after applying the logit transformation, the study-specific sensitivities and specificities are assumed to follow a bivariate normal distribution, consistent with the structure used in the BLMM framework.

The marginal likelihood function of the BGLMM is given by:
\begin{equation}
\label{Eq_6}
L(\boldsymbol{\mu}, \boldsymbol{\Sigma} \mid \boldsymbol{\mu_i}) = \int_{\mathbb{R}^2} \prod_{i=1}^k \text{Binomial}(n_{1i}, Se_i)\text{Binomial}(n_{2i}, Sp_i)\phi_2(\boldsymbol{\mu}_i\ ;\boldsymbol{\mu},\boldsymbol{\Sigma})d\boldsymbol{\mu}_i\tag{3.8}.
\end{equation}
The marginal likelihood function in \ref{Eq_6} does not have a closed-form solution. Consequently, numerical integration methods, including the Laplace Approximation (LA),\cite{pinheiro1995approximations, liu1994note} Iteratively Reweighted Least Squares (IRLS),\cite{jorgensen2006iteratively} and Adaptive Gauss--Hermite Quadrature (AGHQ),\cite{pinheiro1995approximations} are commonly used to approximate the likelihood.\cite{zhao2024evaluation} In this manuscript, the BGLMM was estimated sequentially using AGHQ, LA, and IRLS, with each successive algorithm applied only if the preceding method failed to converge.

\subsection{The Proposed Models}
In this section, we discuss the proposed parametric and nonparametric bootstrap methods. A general review of the bootstrap approach, including methods for estimating standard errors, can be found in Efron (1982)\cite{efron_5_1982} and Efron and Gong (1983). \cite{efron_leisurely_1983}

\subsubsection{The Nonparametric Bootstrap}

The proposed NPB approach estimates the pooled $Se$ and $Sp$ without imposing parametric assumptions on the latent random-effects distribution. Suppose that for \(i = 1, \dots, k\) studies, the observed diagnostic test outcomes are given by \(TP_i\), \(FP_i\), \(TN_i\), and \(FN_i\). The NPB procedure is implemented as follows.

\begin{itemize}

\item \textbf{Step 1:} Draw a bootstrap sample of size \(k\) studies with replacement from the original set of studies. Each sampled study contributes its complete \(2\times2\) contingency table counts \((TP_i, FP_i, TN_i, FN_i)\) to the bootstrap dataset. Consequently, some studies may appear multiple times in a bootstrap sample, while others may not appear.

\item \textbf{Step 2:} For each bootstrap replicate \(b = 1, \dots, B\), compute the study-specific diseased and non-diseased sample sizes:
\[
n_{1i}^{(b)} = TP_i^{(b)} + FN_i^{(b)}, \qquad
n_{2i}^{(b)} = TN_i^{(b)} + FP_i^{(b)}.
\]

A Haldane--Anscombe correction\cite{haldane1956estimation,anscombe1956estimating} was applied only when zero cells were present in the \(2\times2\) table.

The bootstrap estimates of study-specific sensitivity and specificity are then computed as
\[
\widehat{Se}_i^{(b)} = \frac{TP_i^{(b)}}{n_{1i}^{(b)}}, \qquad
\widehat{Sp}_i^{(b)} = \frac{TN_i^{(b)}}{n_{2i}^{(b)}}.
\]

The corresponding logit-transformed estimates are
\(
\boldsymbol{y}_i^{(b)} =
\left(\operatorname{logit}\left(\widehat{Se}_i^{(b)}\right),
\operatorname{logit}\left(\widehat{Sp}_i^{(b)}\right)\right)^{\top}.
\)
The bootstrapped within-study variances were approximated using the delta method:
\(
\psi_{1i}^{(b)} \approx
\frac{1}{n_{1i}^{(b)}\widehat{Se}_i^{(b)}\left(1-\widehat{Se}_i^{(b)}\right)}
\) and 
\(
\psi_{2i}^{(b)} \approx
\frac{1}{n_{2i}^{(b)}\widehat{Sp}_i^{(b)}\left(1-\widehat{Sp}_i^{(b)}\right)}.
\)
Therefore, the within-study covariance matrix is 
\(
\boldsymbol{\Psi}_i^{(b)} =
\begin{pmatrix}
\psi_{1i}^{(b)} & 0 \\
0 & \psi_{2i}^{(b)}
\end{pmatrix}.
\)

The between-study variance-covariance matrix for bootstrap replicate \(b\) is estimated using the sample covariance matrix of the bootstrap vectors \(\boldsymbol{y}_i^{(b)}\):
\[
\widehat{\boldsymbol{\Sigma}}^{(b)}
=
\frac{1}{k-1}
\sum_{i=1}^{k}
\left(
\boldsymbol{y}_i^{(b)}
-
\overline{\boldsymbol{y}}^{(b)}
\right)
\left(
\boldsymbol{y}_i^{(b)}
-
\overline{\boldsymbol{y}}^{(b)}
\right)^\top,
\]
where
\[
\overline{\boldsymbol{y}}^{(b)}
=
\frac{1}{k}
\sum_{i=1}^{k}
\boldsymbol{y}_i^{(b)}.
\]

\item \textbf{Step 3:} For each bootstrap replicate, the pooled logit sensitivity and specificity are estimated using weighted least squares:
\[
\widehat{\boldsymbol{\mu}}^{(b)}
=
\left(
\sum_{i=1}^{k}
\widehat{\boldsymbol{W}}_i^{(b)}
\right)^{-1}
\left(
\sum_{i=1}^{k}
\widehat{\boldsymbol{W}}_i^{(b)}
\boldsymbol{y}_i^{(b)}
\right),
\]
where
\[
\widehat{\boldsymbol{W}}_i^{(b)}
=
\left(
\widehat{\boldsymbol{\Sigma}}^{(b)}
+
\boldsymbol{\Psi}_i^{(b)}
\right)^{-1}.
\]

Weighted least squares estimation was used to account for the apparent heterogeneity in DTA meta-analyses within- and between-studies. Here \(\widehat{\mu}_1^{(b)}\) and \(\widehat{\mu}_2^{(b)}\) denote the bootstrap replicate's pooled sensitivity and specificity on the logit scale, and \(\widehat{\boldsymbol\Sigma}^{(b)}\) (from Step 2) is retained directly as the between-study covariance estimate for replicate \(b\).

\item \textbf{Step 4:} Steps 1 through 3 are repeated independently \(B\) times to obtain bootstrap replicates of the logit-scale pooled estimates,
\[
\widehat{\mu}_1^{(1)},
\widehat{\mu}_1^{(2)},
\dots,
\widehat{\mu}_1^{(B)},
\qquad
\widehat{\mu}_2^{(1)},
\widehat{\mu}_2^{(2)},
\dots,
\widehat{\mu}_2^{(B)}.
\]
The final pooled point estimates were obtained by averaging on the logit scale and back-transforming once to the probability scale, using the inverse-logit transformation:
\[
\widehat{Se}_{NPB}
=
\operatorname{expit}\!\left(
\frac{1}{B}\sum_{b=1}^{B}\widehat{\mu}_1^{(b)}
\right),
\qquad
\widehat{Sp}_{NPB}
=
\operatorname{expit}\!\left(
\frac{1}{B}\sum_{b=1}^{B}\widehat{\mu}_2^{(b)}
\right),
\]
where \(\operatorname{expit}(x)=1/\{1+e^{-x}\}\). 

95\% CIs for the pooled sensitivity and specificity were likewise constructed on the logit scale, using the empirical \(2.5^{th}\) and \(97.5^{th}\) percentiles of the logit-scale bootstrap replicates, and then back-transformed to the probability scale:
\[
CI_{Se}
=
\left[
\operatorname{expit}\!\left(Q_{0.025}\left(\left\{\widehat{\mu}_1^{(b)}\right\}_{b=1}^{B}\right)\right),
\ \
\operatorname{expit}\!\left(Q_{0.975}\left(\left\{\widehat{\mu}_1^{(b)}\right\}_{b=1}^{B}\right)\right)
\right],
\]
\[
CI_{Sp}
=
\left[
\operatorname{expit}\!\left(Q_{0.025}\left(\left\{\widehat{\mu}_2^{(b)}\right\}_{b=1}^{B}\right)\right),
\ \
\operatorname{expit}\!\left(Q_{0.975}\left(\left\{\widehat{\mu}_2^{(b)}\right\}_{b=1}^{B}\right)\right)
\right],
\]
where \(Q_p(\cdot)\) denotes the empirical \(p\)-th quantile of the bootstrap replicates. The between-study variance-covariance components were reported as simple averages of \(\widehat{\boldsymbol\Sigma}^{(b)}\) across the \(B\) bootstrap replicates,
\[
\overline{\sigma}_{11}
=
\frac{1}{B}\sum_{b=1}^{B}\widehat{\Sigma}_{11}^{(b)},
\qquad
\overline{\sigma}_{22}
=
\frac{1}{B}\sum_{b=1}^{B}\widehat{\Sigma}_{22}^{(b)},
\qquad
\overline{\sigma}_{12}
=
\frac{1}{B}\sum_{b=1}^{B}\overline{\Sigma}_{12}^{(b)}.
\]
\end{itemize}

\subsubsection{Parametric Bootstrap}
\label{section:3.4.2}
The proposed PB approach generates bootstrap datasets assuming a parametric random-effects distribution. In contrast to the nonparametric bootstrap approach, the PB methods assume a parametric distribution for the latent study-specific random effects and refit the BGLMM to each bootstrap sample. For each bootstrap replicate \(b = 1, \dots, B\), the procedure is implemented as follows.

\begin{itemize}

\item \textbf{Step 1:} Fit the BGLMM to the observed data to obtain estimates of the pooled logit sensitivity and specificity,
\[
\widehat{\boldsymbol{\mu}}
=
\begin{pmatrix}
\widehat{\mu}_1 \\
\widehat{\mu}_2
\end{pmatrix},
\]
and the estimated between-study heterogeneity,
\[
\widehat{\boldsymbol{\Sigma}}
=
\begin{pmatrix}
\widehat{\sigma}_1^2 & \widehat{\sigma}_{12} \\
\widehat{\sigma}_{12} & \widehat{\sigma}_2^2
\end{pmatrix}.
\]

\item \textbf{Step 2:} Bootstrapped study-specific latent effects are generated from one of three assumed bivariate random-effects distributions for the latent logit-scale sensitivity and specificity, matching the distributional assumption under evaluation:
\[
\boldsymbol{y}_i^{(b)}= \left(y_{1i}^{(b)}, y_{2i}^{(b)}\right)^{\top}
\sim
\begin{cases}
N_2\left(\widehat{\boldsymbol{\mu}},\ \widehat{\boldsymbol{\Sigma}}\right),
 & \text{bivariate normal,} \\[4pt]
t_2\left(\nu=3,\ \widehat{\boldsymbol{\mu}},\ \widehat{\boldsymbol{\Sigma}}\right),
 & \text{bivariate Student-}t\text{ (3 df),} \\[4pt]
L_2\left(\widehat{\boldsymbol{\mu}},\ \widehat{\boldsymbol{\Sigma}}\right),
 & \text{bivariate (symmetric) Laplace,}
\end{cases}
\]
where \(t_2(\nu=3,\cdot,\cdot)\) denotes the bivariate Student-\(t\) distribution with 3 degrees of freedom and scale matrix \(\widehat{\boldsymbol\Sigma}\), and \(L_2(\cdot,\cdot)\) denotes the bivariate symmetric Laplace distribution with location \(\widehat{\boldsymbol\mu}\) and scale matrix \(\widehat{\boldsymbol\Sigma}\), both of which which are obtained in Step 2. This allows the sensitivity of the PB and BCPB estimators to the assumed random-effects distribution to be assessed directly within the simulation study.

\item \textbf{Step 3:} The simulated latent effects are transformed back to the probability scale using the inverse-logit transformation:
\[
Se_i^{(b)}
=
\frac{
\exp\left(y_{1i}^{(b)}\right)
}{
1 + \exp\left(y_{1i}^{(b)}\right)
},
\qquad
Sp_i^{(b)}
=
\frac{
\exp\left(y_{2i}^{(b)}\right)
}{
1 + \exp\left(y_{2i}^{(b)}\right)
}.
\]

\item \textbf{Step 4:} The observed diseased and non-diseased sample sizes for each study were preserved throughout the bootstrap procedure. Conditional on these fixed sample sizes, bootstrapped \(2\times2\) table counts were generated from independent binomial distributions:
\[
TP_i^{(b)}
\sim
\mathrm{Binomial}\left(n_{1i}, Se_i^{(b)}\right),
\qquad
FN_i^{(b)}
=
n_{1i} - TP_i^{(b)},
\]
\[
TN_i^{(b)}
\sim
\mathrm{Binomial}(n_{2i}, Sp_i^{(b)}),
\qquad
FP_i^{(b)}
=
n_{2i} - TN_i^{(b)}.
\]

\item \textbf{Step 5:} Fit the BGLMM to each bootstrap dataset to obtain bootstrap estimates of the pooled logit sensitivity and specificity,
\[
\widehat{\boldsymbol{\mu}}^{(b)}=
\begin{pmatrix}
\widehat{\mu_1}^{(b)}\\
\widehat{\mu}_2^{(b)}
\end{pmatrix},
\]
and the between-study covariance matrix,
\[
\widehat{\boldsymbol{\Sigma}}^{(b)}.
\]

\item \textbf{Step 6:} Steps 1 through 5 are repeated independently \(B\) times to obtain bootstrap replicates
\[
\widehat{\mu_1}^{(1)},
\widehat{\mu_1}^{(2)},
\dots,
\widehat{\mu_1}^{(B)},
\qquad
\widehat{\mu_2}^{(1)},
\widehat{\mu_2}^{(2)},
\dots,
\widehat{\mu_2}^{(B)}.
\]

As with the non-parametric bootstrap (NPB), the parametric bootstrap (PB) point estimates and CIs are computed on the logit scale and subsequently back-transformed to the probability scale using the inverse-logit function. Specifically, the pooled sensitivity and specificity are estimated by first averaging the bootstrap estimates on the logit scale and then applying a single inverse-logit transformation:
\[
\widehat{Se}_{PB}
=
\operatorname{expit}\!\left(
\frac{1}{B}\sum_{b=1}^{B}\widehat{\mu}_1^{(b)}
\right),
\qquad
\widehat{Sp}_{PB}
=
\operatorname{expit}\!\left(
\frac{1}{B}\sum_{b=1}^{B}\widehat{\mu}_2^{(b)}
\right).
\]

The corresponding 95\% CIs were constructed using the percentile bootstrap method:
\[
CI_{Se}
=
\left[
\operatorname{expit}\!\left(Q_{0.025}\left(\left\{\widehat{\mu}_1^{(b)}\right\}_{b=1}^{B}\right)\right),
\ \
\operatorname{expit}\!\left(Q_{0.975}\left(\left\{\widehat{\mu}_1^{(b)}\right\}_{b=1}^{B}\right)\right)
\right],
\]
\[
CI_{Sp}
=
\left[
\operatorname{expit}\!\left(Q_{0.025}\left(\left\{\widehat{\mu}_2^{(b)}\right\}_{b=1}^{B}\right)\right),
\ \
\operatorname{expit}\!\left(Q_{0.975}\left(\left\{\widehat{\mu}_2^{(b)}\right\}_{b=1}^{B}\right)\right)
\right],
\]
where \(Q_p(\cdot)\) denotes the empirical \(p\)-th quantile of the bootstrap replicates. The between-study variance-covariance components are reported as simple averages of \(\widehat{\boldsymbol\Sigma}^{(b)}\) across the \(B\) bootstrap replicates, analogous to the NPB.
\end{itemize}
\subsubsection{Bias-Corrected Parametric Bootstrap}
Bias-corrected parametric bootstrap methods have become popular tools for assessing estimation uncertainty. Bootstrap estimators may also exhibit finite-sample bias, making it important to account for this bias when conducting statistical inference.\cite{steck2003bias} Motivated by this consideration, we propose a BCPB procedure that reduces the finite-sample bias of the PB estimator using Efron's reflection principle.\cite{tibshirani1993introduction}
Let
\(
\widehat{\boldsymbol{\mu}}
\)
denote the logit of the original BGLMM estimates obtained from the observed data (Step 1 of the PB procedure), and let
\[
\overline{\boldsymbol{\mu}}
=
\frac{1}{B}\sum_{b=1}^{B}\hat{\boldsymbol{\mu}}^{(b)}
\]
denote the mean of the logit-transformed PB bootstrap replicates from Step 6 of the PB procedure.
The bias-corrected bootstrap replicates on the logit-scale are obtained by reflecting each bootstrap estimate about the original BGLMM estimate as
\[
\widehat{\mu}_{BC}^{(b)}
=
2\widehat{\mu}
-
\widehat{\mu}^{(b)},
\qquad b=1,\ldots,B.
\]

Then, the bias-corrected point estimates \(\left(\widehat{\boldsymbol{\mu}}_{BC}\right)\) on the logit scale is given by
\[
\widehat{\boldsymbol{\mu}}_{BC}
=
2\,\widehat{\boldsymbol{\mu}}
-
\overline{\boldsymbol{\mu}}
\]
The same reflection principle is applied to the between-study variance-covariance matrix. Let \(\widehat{\sigma}_{1}^{2}, \ \widehat{\sigma}_{2}^{2}, \ \widehat{\sigma}_{12}\) denote the entries of \(\widehat{\boldsymbol\Sigma}\) from the original BGLMM fit (Step 1 of the PB procedure), and let \(\overline{\sigma}_{1}^{2},\ \overline{\sigma}_{2}^{2}, \ \overline{\sigma}_{12}\) denote the corresponding averages of \(\widehat{\boldsymbol\Sigma}^{(b)}\) across the \(B\) bootstrap replicates (Steps 5 and 6 of the PB procedure). The bias-corrected between-study variance-covariance estimates are
\[
\widehat{\sigma}_{1}^{2(BC)}
=
\max\!\big(0,\ 2\widehat{\sigma}_{1}^{2} - \overline{\sigma}_{1}^{2}\big),
\qquad
\widehat{\sigma}_{2}^{2(BC)}
=
\max\!\big(0,\ 2\widehat{\sigma}_{2}^{2} - \overline{\sigma}_{2}^{2}\big), \text{ and } 
\widehat{\sigma}_{12}^{(BC)}
=
2\widehat{\sigma}_{12} - \overline{\sigma}_{12},
\]
where the variances are truncated at zero to respect the non-negativity constraint on variances, and the covariance component is left untruncated, since it may legitimately be negative.

The bias-corrected Wald CI for the true logit-transformed \(Se\) and \(Sp\) are constructed using the BCPB estimate and its bootstrap standard error. Specifically, the $100(1-\alpha)\%$ CI for \(\text{logit}(Se)\) is given by 
\[
\widehat{\mu}_{BC,1}
\pm
z_{1-\alpha/2}\,
\widehat{SE}_{\widehat{\mu}_{BC,1}},
\]
where \(z_{1-\alpha/2}=\Phi^{-1}(1-\alpha/2)\) is the \((1-\alpha/2)\)-quantile of the standard normal distribution and
\[
\widehat{SE}_{\widehat{\mu}_{BC,1}}
=
\sqrt{
\frac{1}{B-1}
\sum_{b=1}^{B}
\left(
\widehat{\mu}_{BC,1}^{(b)}
-
\widehat{\mu}_{BC,1}
\right)^2
},
\]
with
\[
\widehat{\mu}_{BC,1}
=
\frac{1}{B}
\sum_{b=1}^{B}
\widehat{\mu}_{BC,1}^{(b)}.
\]
The confidence limits are subsequently back-transformed to the probability scale using the inverse-logit function. The confidence interval for \(Sp\) is constructed analogously using
\[
\widehat{\mu}_{BC,2}
\pm
z_{1-\alpha/2}\,
\widehat{SE}_{\widehat{\mu}_{BC,2}},
\]
where
\[
\widehat{SE}_{\widehat{\mu}_{BC,2}}
=
\sqrt{
\frac{1}{B-1}
\sum_{b=1}^{B}
\left(
\widehat{\mu}_{BC,2}^{(b)}
-
\widehat{\mu}_{BC,2}
\right)^2
},
\]
and
\[
\widehat{\mu}_{BC,2}
=
\frac{1}{B}
\sum_{b=1}^{B}
\widehat{\mu}_{BC,2}^{(b)}.
\]

In this study, bootstrap procedures were implemented using \(B = 500\) bootstrap replications. This number was selected to balance computational efficiency and estimation precision, given the large number of simulation scenarios and replications considered in the manuscript.
%\subsection{}

\section{Simulation Study}
\label{s_5}
\subsection{Simulation Study Design}\label{s_4}
By mimicking the characteristics of the TEE data\cite{van_zaane_meta-analysis_2008}, EUS data\cite{mocellin_diagnostic_nodate} and Zhao \textit{et al.} (2024)\cite{zhao2024evaluation}, the simulation study has been designed to compare the performance of the proposed bootstrap methods (NPB, PB and BCPB) with the standard models (BLMM and BGLMM) in terms of several empirical performance measures:\cite{burton_design_2006,morris_using_2019} bias, 95\% CI width, root mean squared error (RMSE), and the 95\% coverage probability (CP). The true parameters in setting 1 of Table \ref{Table_2} were based on the TEE data\cite{van_zaane_meta-analysis_2008}. For setting 2, the true parameters were drawn from the EUS data\cite{mocellin_diagnostic_nodate}. Lastly, in setting 3, the true parameters were based on the simulation design of Zhao \textit{et al.} (2024)\cite{zhao2024evaluation}, who extracted data from the Cochrane Database of Systematic Reviews and selected 64 reviews containing 393 meta-analyses, to generate 1000 meta-analyses for each simulation setting. Accordingly, we considered a total of $3^4\times4=324$ scenarios in our simulation study, as depicted in Table \ref{Table_2}, by combining the between-study covariance parameter ($\sigma_{12} $) with the between-study variances $(\sigma_1^2, \sigma_2^2)$. 

\begin{table}[ht]
\centering
\caption{ True parameter settings for the simulation study}
\label{Table_2}
\begin{tabular}{lcccc}
      \toprule
\textbf{Parameter} & \textbf{Setting 1} & \textbf{Setting 2} & \textbf{Setting 3} & \textbf{Setting 4}  \\
\toprule
$(Se, Sp)$   & (0.5, 0.6)     & (0.7, 0.8)   & (0.95, 0.99 )  & -   \\
RE distribution   &   Bivariate student t   & Bivariate Normal   & Bivariate Laplace  & -   \\
$(n_1, n_2)$           & (20, 50)      & (50, 100)     & (100,200)      & -     \\
$(\sigma_1^2, \sigma_2^2)$ & (0.90, 0.57) & (0.54, 0.95)     & (1.59, 1.83)    & -  \\
$\sigma_{12}$         & -0.22          & -0.56        & -0.70       & -         \\
$k$                   &       5      &     20        &  40        &  100    \\
    \bottomrule
\end{tabular}
\end{table}

\subsubsection{Data Generation}
For the simulation studies, DTA data were generated following the procedure outlined below:\cite{negeri_skew-normal_2020}
\begin{itemize}
   \item \textbf{step 1:} The study-specific logit($Se_i$) and logit($Sp_i$) pairs were simulated from three assumed random-effects distributions: the bivariate normal, bivariate Student-$t$, and bivariate Laplace distributions, with specified mean ($\boldsymbol{\mu}$), covariance ($\boldsymbol{\Sigma}$), and other data characteristics described in Table \ref{Table_2}.
    \item \textbf{Step 2:} The generated logit($Se_i$) and logit($Sp_i$) values in Step 1 were transformed to the probability scale using the inverse-logit function to obtain the corresponding true study-specific sensitivity ($Se_i$) and specificity ($Sp_i$).
   \item \textbf{Step 3:} The study-specific frequencies ($TP_i, \ FN_i,\ TN_i, $ and $FP_i$) were generated from the binomial distribution using the proportions simulated in Step 2 and the study-specific sample sizes ($n_{i1}, \ n_{i2}$) independently generated from Poisson distribution with means ($n_1,n_2$) from Table~\ref{Table_2}. 
\end{itemize}

We used \texttt{R} statistical language\cite{R2025} version 4.5.1 (2025-06-13) and RStudio\cite{R2026} version 4.6.0 (2026-04-24 ucrt) for data simulation and statistical analyses.  The \texttt{mada}\cite{doebler_mada_2025} R package was used to fit the BLMM using the \texttt{reitsma()} function via the maximum likelihood (ML) estimation method.

Data from the bivariate normal and bivariate Student's $t$ distributions were simulated using the \texttt{mvrnorm()} and \texttt{rmvt()} functions, respectively, from the \texttt{mvtnorm}\cite{genz2021package} R package, whereas data from the bivariate Laplace distribution were simulated using the \texttt{rmvl()} function from the \texttt{LaplacesDemon}\cite{statisticat2013laplacesdemon} R package. We wrote our own \texttt{R} functions to fit the proposed bootstrap approaches, whereas the BGLMM was by default fitted using the Adaptive Gauss–Hermite Quadrature (AGHQ) algorithm with the \texttt{mixed\_model()} function from the \texttt{GLMMadaptive}\cite{Rizopoulos2023} \texttt{R} package, setting the number of quadrature points (\texttt{nAGQ}) to 5; models that failed to converge were sequentially refitted using the Laplace approximation (LA) and Iteratively Reweighted Least Squares (IRLS) algorithm with the \texttt{glmer()} function from the \texttt{lme4}\cite{bates2015fitting} \texttt{R} package, setting the \texttt{nAGQ} to 1 and 0, respectively. To ensure reproducibility, the simulations were run 1000 times using different seeds for each replication.

\subsection{Simulation Study Results}
The simulation results reported herein correspond to scenarios with small participants in each group (i.e., $n_1=20, n_2=50$ in the diseased and non-diseased group, respectively),  large between-study variances for $\text{logit}(Se)$ and $\text{logit}(Sp)$, and a small negative covariance between $\text{logit}(Se)$ and $\text{logit}(Sp)$ across the three distributions.

Because the relative performance and ranking of the competing methods remained largely unchanged across all three random-effects distributions, only results and conclusions drawn from the bivariate normal scenarios are presented in detail, as they are representative of the broader simulation study.

\begin{figure}[ht]
    \centering
    \includegraphics[width=\linewidth]{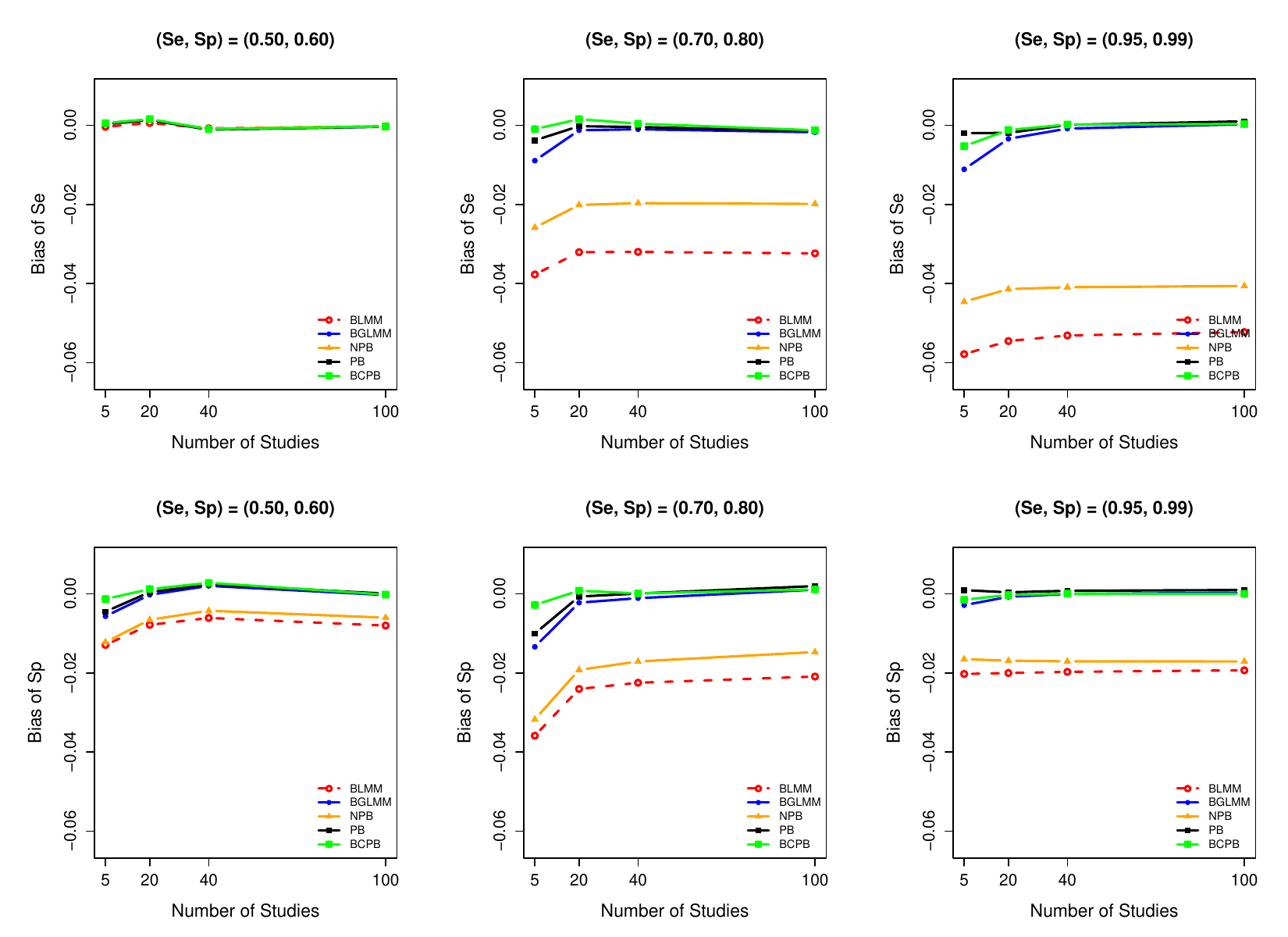}
    \caption{Bias of $Se$ and $Sp$ for different combinations of studies and 
    true ($Se$, $Sp$) pairs when $\sigma_1^2=1.59$, $\sigma_{12}=-0.22$, 
    $\sigma_2^2=1.83$, $n_1=20$, $n_2=50$ and random effects were generated 
    from the normal distribution.}
    \label{fig_4.1}
\end{figure}

Figure \ref{fig_4.1} presents the bias of the overall $Se$ and $Sp$ estimators for the BLMM, BGLMM, NPB, PB and BCPB approaches across different numbers of studies and diagnostic accuracy settings when random effects were generated from the bivariate normal distribution. Overall, most methods produced negatively biased estimates across several simulation settings, although the magnitude and direction of bias varied considerably between methods and scenarios. For the low diagnostic accuracy setting \((Se, Sp) = (0.50, 0.60)\), all five methods produced negligible bias for sensitivity across all number of studies. For specificity, the BLMM and NPB exhibited persistent slight negative bias across all number of studies, whereas the BGLMM and PB showed noticeable improvement as the number of studies increased, with their bias approaching zero. Notably, the proposed BCPB approach produced specificity estimates with bias close to zero even when only five studies were available and maintained this performance across all values of $k$. 

As the true sensitivity and specificity increased to \((0.70, 0.80)\) and \((0.95, 0.99)\), clearer differences among the competing approaches emerged. The BLMM method consistently exhibited the highest negative bias for both sensitivity and specificity estimates, and this bias remained relatively stable as the number of studies increased. The BGLMM showed substantially reduced bias, approaching zero as the number of studies increased. The NPB approach exhibited a persistent negative bias, particularly under high diagnostic accuracy settings \((0.95, 0.99)\), whereas the PB approach produced estimates that remained close to the true values across all study sizes. The proposed BCPB approach likewise exhibited negligible bias throughout and generally provided estimates closest to the true values across the simulation settings.

The proposed NPB and PB approaches demonstrated intermediate performance, with bias values generally lower than those obtained from the BLMM method. In high diagnostic accuracy settings, the PB approach showed improved bias performance relative to the NPB approach for both sensitivity and specificity estimates. Nevertheless, the BCPB approach achieved the lowest absolute bias and remained closest to zero across most study sizes, demonstrating the effectiveness of the bootstrap bias correction method.

When the within-study sample sizes were relatively large (\(n_1=100\), \(n_2=200\); Figure 26 in Appendix B), the magnitude of bias was generally small across the competing methods. Nevertheless, differences among methods were most apparent when the number of studies was small (\(k=5\)). For \((Se, Sp)=(0.50, 0.60)\), all five methods produced bias values close to zero, particularly for sensitivity. As the diagnostic accuracy increased to \((Se, Sp)=(0.70, 0.80)\) and \((0.95, 0.99)\), clearer differences emerged. The BLMM exhibited the greatest negative bias, while the NPB also showed noticeable negative bias in several settings. In contrast, the proposed BCPB approach consistently produced estimates with bias closest to zero for both sensitivity and specificity, including when only five studies were available. The BGLMM and PB also performed well, with their bias decreasing rapidly as the number of studies increased. Overall, the BCPB demonstrated excellent bias performance across all three diagnostic accuracy settings and study sizes, with its advantage most evident in small meta-analyses and in moderate-to-high diagnostic accuracy settings.

\begin{figure}[ht]
    \centering
    \includegraphics[width=\linewidth]{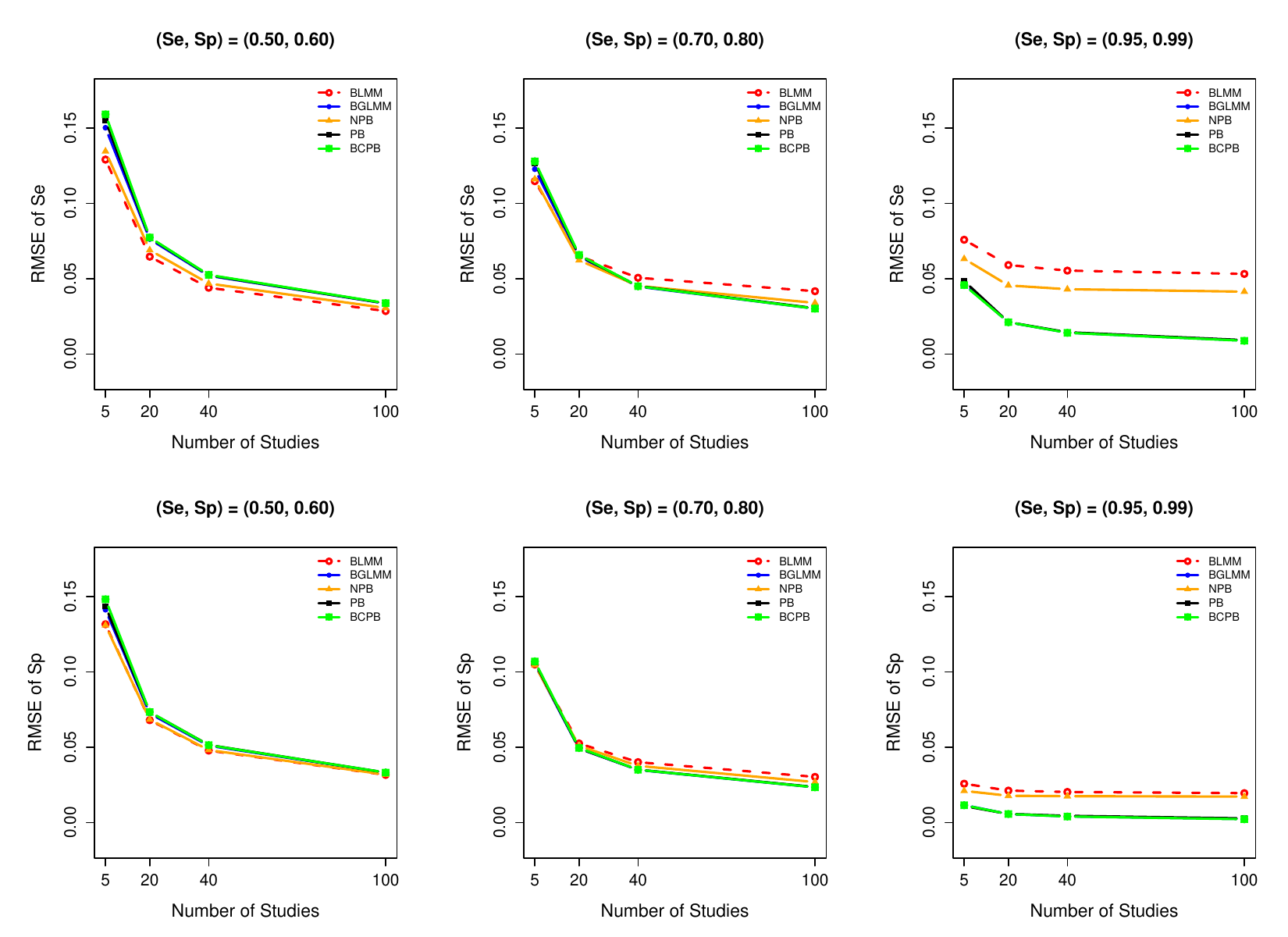}
    \caption{RMSE of $Se$ and $Sp$ for different combinations of studies and 
    true ($Se$, $Sp$) pairs when $\sigma_1^2=1.59$, $\sigma_{12}=-0.22$, 
    $\sigma_2^2=1.83$, $n_1=20$, $n_2=50$ and random effects were generated 
    from the normal distribution.}
    \label{fig_4.2}
\end{figure}

Figure~\ref{fig_4.2} presents the RMSE of estimates for the BLMM, BGLMM, NPB, PB and BCPB approaches under the smaller within-study sample size setting ($n_1 = 20, n_2 = 50$). Overall, the RMSE values decreased as the number of studies increased, indicating improved estimation precision for all competing methods.

For the moderate diagnostic accuracy setting \((Se, Sp)=(0.50, 0.60)\), all five methods exhibited very similar RMSE patterns for both sensitivity and specificity. The RMSE decreased substantially as the number of studies increased from \(k=5\) to \(k=100\), and the differences among methods became negligible for larger number of studies.

The largest differences among methods were observed in the high diagnostic accuracy setting \((Se, Sp)=(0.95, 0.99)\). Under this setting, the BLMM consistently produced the highest RMSE values for both sensitivity and specificity across all study sizes. The proposed NPB approach reduced the RMSE relative to the BLMM, although its RMSE remained consistently higher than those of the BGLMM, PB, and BCPB approaches. In contrast, the BGLMM, PB, and BCPB approaches all exhibited excellent performance, producing substantially lower RMSE values for both sensitivity and specificity. Among these methods, the proposed BCPB approach consistently achieved the lowest RMSE, although its improvement over the BGLMM and PB approaches was modest. 

\begin{figure}[ht]
    \centering
    \includegraphics[width=\linewidth]{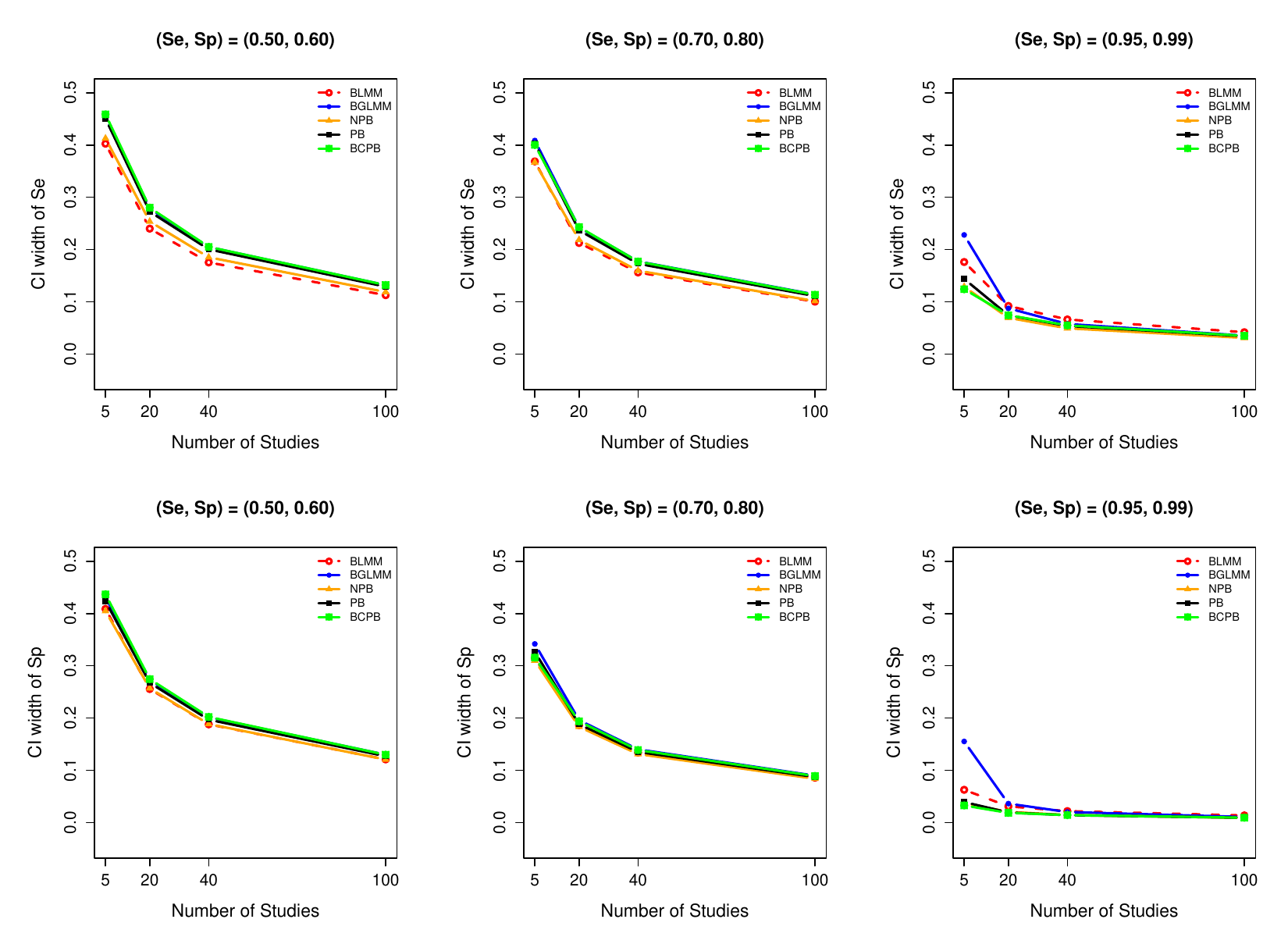}
    \caption{CI widths of $Se$ and $Sp$ for different combinations of studies and true ($Se$, $Sp$) pairs when $\sigma_1^2=1.59$, $\sigma_{12}=-0.22$, 
    $\sigma_2^2=1.83$, $n_1=20$, $n_2=50$ and random effects were generated from the normal distribution.}
    \label{fig_4.4}
\end{figure}

In terms of average 95\% CI width (Figure~\ref{fig_4.4}), all five methods performed similarly for the \((Se, Sp)=(0.50, 0.60)\) and \((0.70, 0.80)\) settings, with CI widths decreasing as the number of studies increased. The BCPB, PB, and NPB approaches produced nearly identical interval widths across these settings.

Under the high diagnostic accuracy setting \((Se, Sp)=(0.95, 0.99)\), more noticeable differences among the methods were observed. The proposed BCPB approach consistently produced the narrowest average CIs for both sensitivity and specificity across all study sizes. The PB and NPB approaches also yielded relatively narrow CIs, whereas the BGLMM produced substantially wider CIs, particularly when $k=5$. As the number of studies increased, the CI widths decreased for all methods and became increasingly similar.

\begin{figure}[ht]
    \centering
    \includegraphics[width=\linewidth]{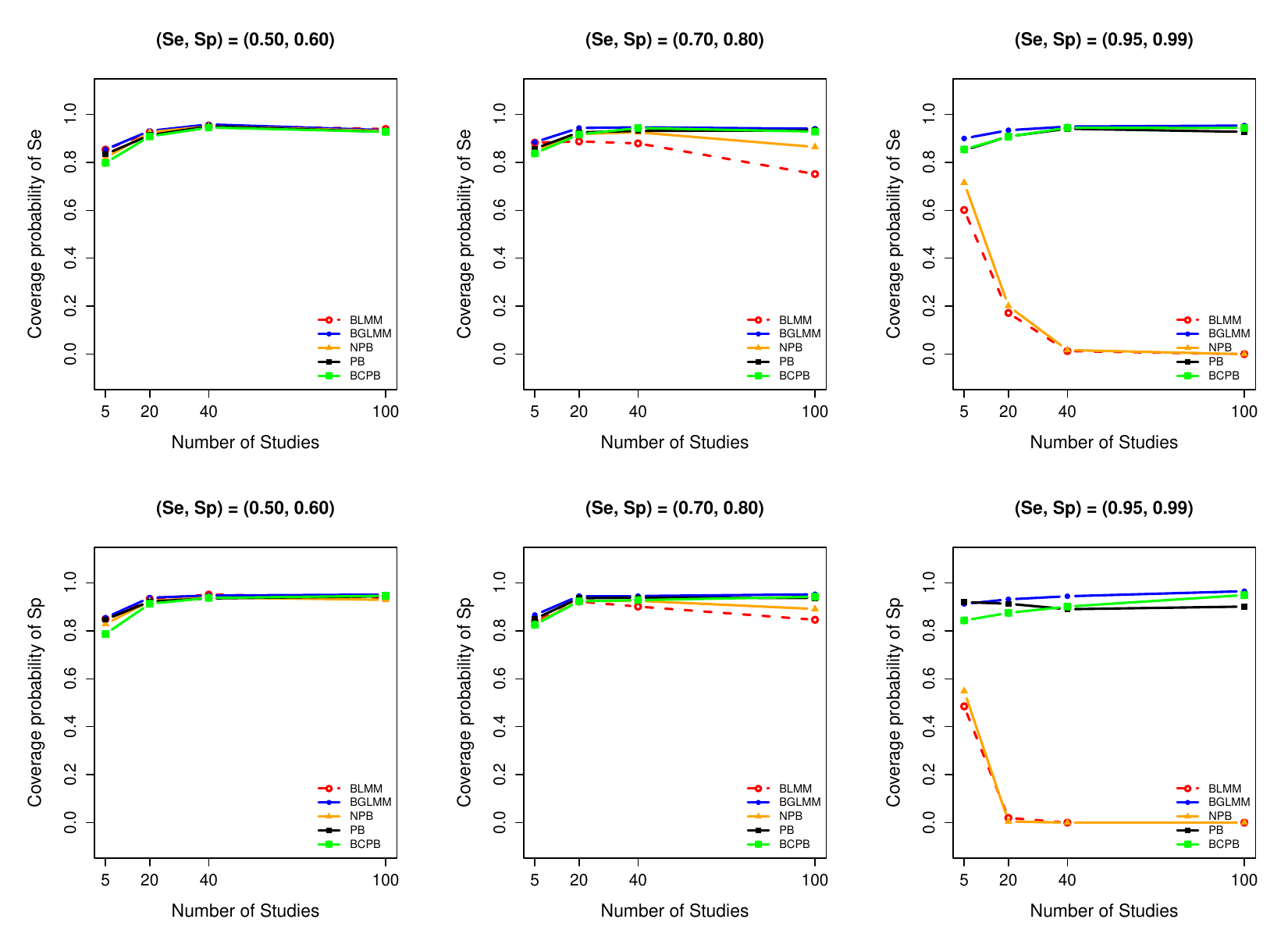}
    \caption{Coverage probability of $Se$ and $Sp$ for different combinations of studies and true ($Se$, $Sp$) pairs when $\sigma_1^2=1.59$, $\sigma_{12}=-0.22$, 
    $\sigma_2^2=1.83$, $n_1=20$, $n_2=50$ and random effects were generated from the normal distribution.}
    \label{fig_4.3}
\end{figure}

Regarding the empirical 95\% CP (Figure~\ref{fig_4.3}), all five methods achieved CPs close to the nominal 95\% level when the true \((Se, Sp)\) pair was \((0.50, 0.60)\), with CP generally improving as the number of studies increased. For the moderate diagnostic accuracy setting \((0.70, 0.80)\), the BGLMM, PB, and BCPB approaches produced very similar empirical CPs and remained close to the nominal 95\% level across moderate and large numbers of studies. The NPB approach exhibited somewhat lower coverage, whereas the BLMM showed a more pronounced decline in CP as the number of studies increased, particularly for sensitivity.

The largest differences among methods were observed in the high diagnostic accuracy setting \((0.95, 0.99)\). In this setting, the BLMM and NPB approaches exhibited substantial undercoverage, with empirical CPs decreasing sharply as the number of studies increased, approaching zero in some scenarios. In contrast, the BGLMM, PB, and BCPB approaches maintained relatively high and stable CPs for both sensitivity and specificity across the study sizes considered. Among these methods, the BGLMM generally produced the highest empirical CPs, while the PB and BCPB approaches also remained close to the nominal 95\% level.

The corresponding empirical CPs for the larger within-study sample size setting (\(n_1=100,\ n_2=200\)) are presented in Figure~26 in Appendix~D of the ancillary files. Overall, increasing the within-study sample sizes substantially improved the coverage performance of all competing approaches. For the \((Se, Sp)=(0.50, 0.60)\) and \((0.70, 0.80)\) settings, all five methods achieved empirical CPs close to the nominal 95\% level across nearly all study sizes. Under the high diagnostic accuracy setting, more noticeable differences among the competing methods remained evident. The BCPB, BGLMM, and PB approaches consistently maintained empirical CPs close to the nominal 95\% level for both sensitivity and specificity, with the BGLMM and BCPB exhibiting the most stable coverage across study sizes. In contrast, the BLMM exhibited substantial undercoverage, particularly as the number of studies increased. The NPB approach also showed declining empirical CPs under this setting, especially for specificity estimation, though its coverage generally remained higher than that of the BLMM.

\begin{figure}[ht]
    \centering
    \includegraphics[width=\linewidth]{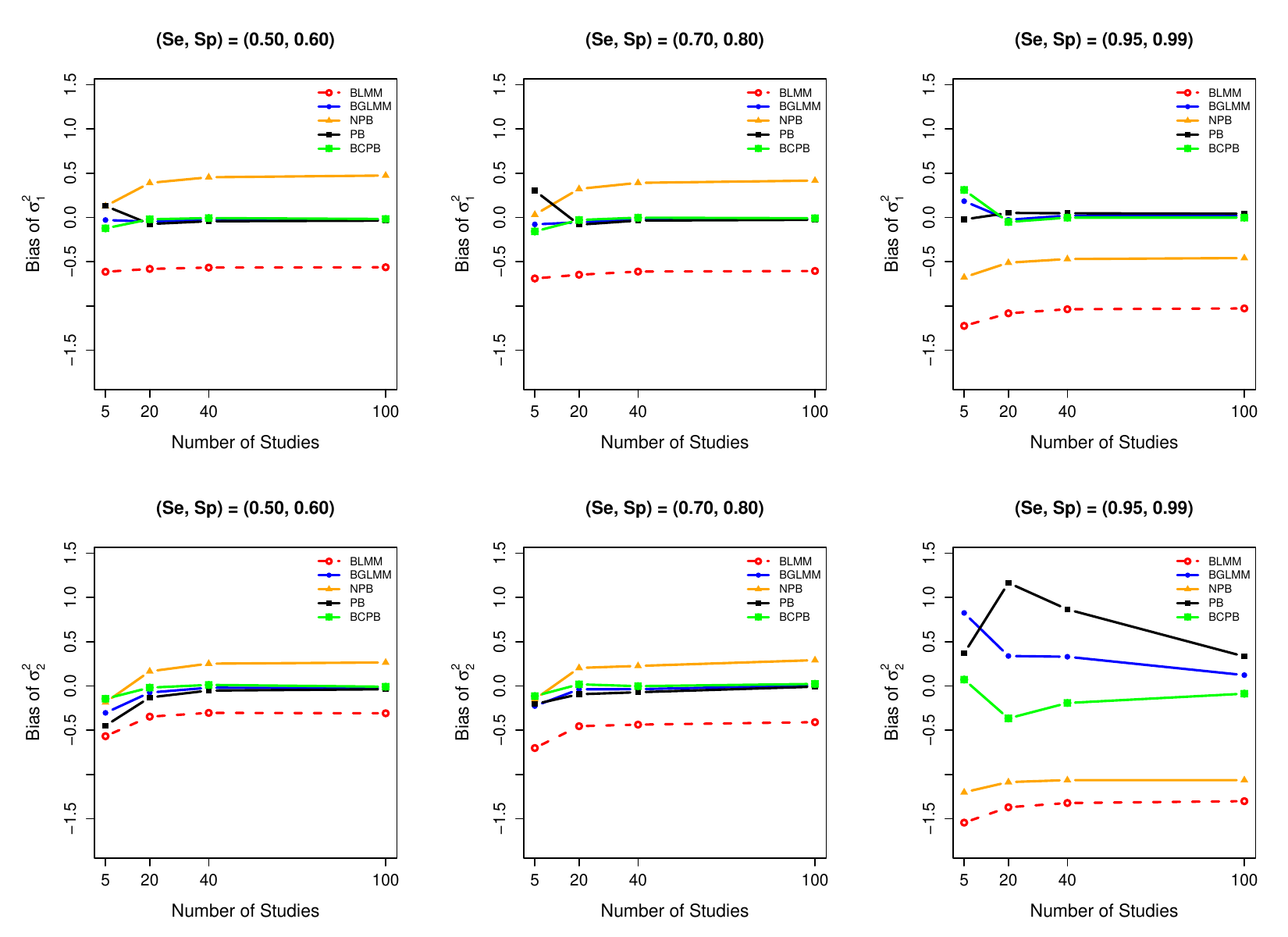}
    \caption{ Bias of $\sigma_1^2$ and $\sigma_2^2 $ for different combinations of studies and true ($Se$, $Sp$) pairs when $\sigma_1^2=1.59$, $\sigma_{12}=-0.22$, 
    $\sigma_2^2=1.83$, $n_1=20$, $n_2=50$ and random effects were generated from the normal distribution.}
    \label{fig_4.5}
\end{figure}

Figure~\ref{fig_4.5} presents the bias of the between-study variance estimates, \(\sigma_1^2\) and \(\sigma_2^2\), under the setting \(n_1=20\), \(n_2=50\), \(\sigma_1^2=1.59\), \(\sigma_2^2=1.83\), and \(\sigma_{12}=-0.22\), with random effects generated from the bivariate normal distribution. For the \((Se, Sp)=(0.50, 0.60)\) and \((0.70, 0.80)\) settings, the BGLMM, PB, and BCPB approaches generally produced bias values close to zero for both variance components across most study sizes. The NPB approach also performed relatively well, although it showed a small positive bias in several settings. In contrast, the BLMM consistently underestimated both \(\sigma_1^2\) and \(\sigma_2^2\), with the negative bias remaining relatively stable as the number of studies increased.

Under the high diagnostic accuracy setting \((Se, Sp)=(0.95, 0.99)\), substantially greater differences among methods were observed. The BLMM approach showed strong negative bias for both variance components across all study sizes. The NPB approach also underestimated both \(\sigma_1^2\) and \(\sigma_2^2\), although its bias was less pronounced than that of the BLMM. The BGLMM produced nearly unbiased estimates of \(\sigma_1^2\) across all study sizes but exhibited positive bias for \(\sigma_2^2\)
 when the number of studies was small $(k=5)$. In contrast, the PB approach substantially overestimated both variance components at $k=5$, with the bias decreasing rapidly as the number of studies increased. The proposed BCPB approach consistently produced variance estimates with bias close to zero in nearly all study sizes.

\begin{figure}[ht]
    \centering
    \includegraphics[width=\linewidth]{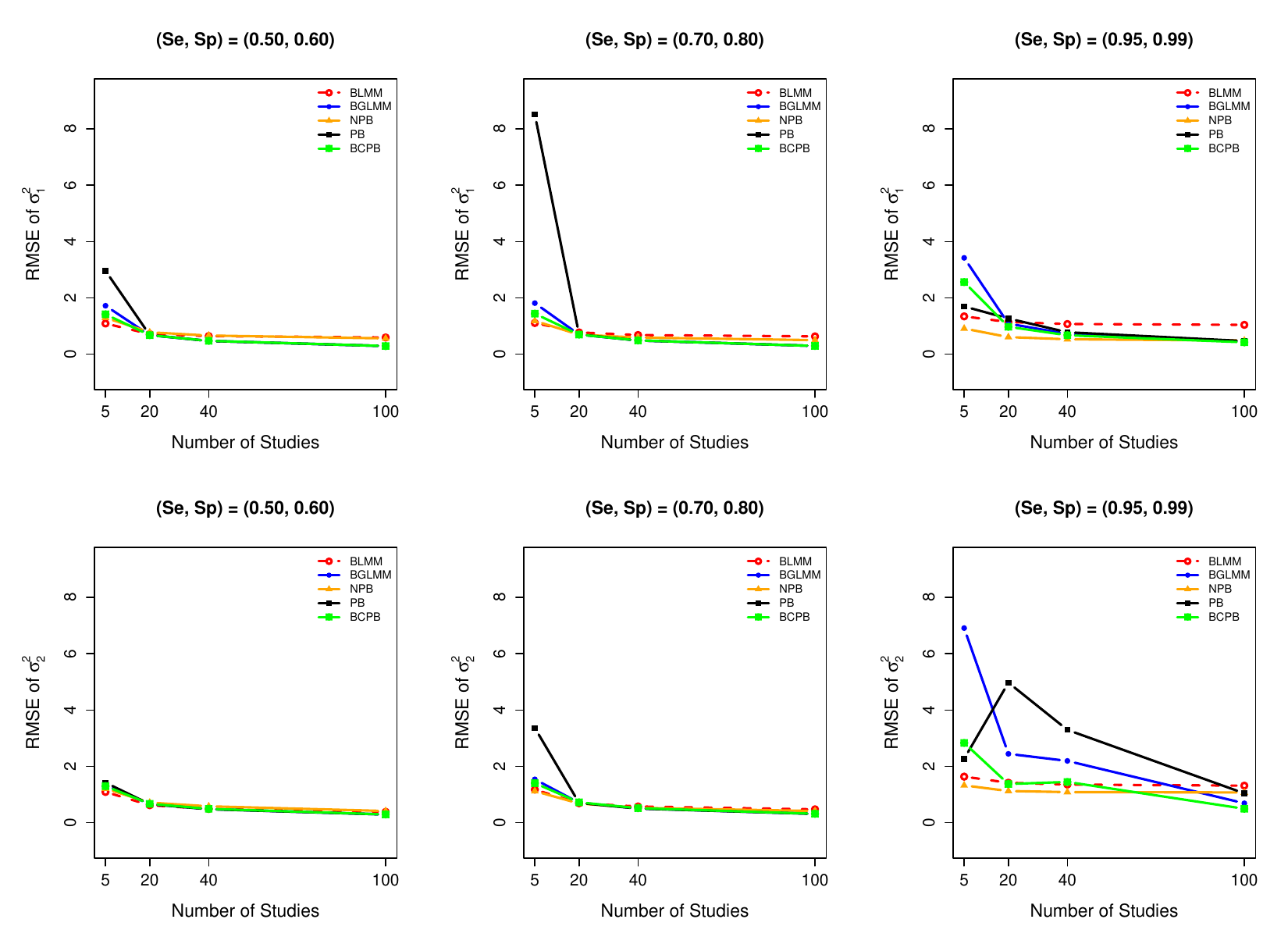}
    \caption{ RMSE of $\sigma_1^2$ and $\sigma_2^2 $ for different combinations of studies and true ($Se$, $Sp$) pairs when $\sigma_1^2=1.59$, $\sigma_{12}=-0.22$, 
    $\sigma_2^2=1.83$, $n_1=20$, $n_2=50$ and random effects were generated from the normal distribution.}
    \label{fig_4.6}
\end{figure}

With regard to the RMSE of the between-study variance estimates (Figure~\ref{fig_4.6}), \(\sigma_1^2\) and \(\sigma_2^2\), under the setting \(n_1=20\), \(n_2=50\), \(\sigma_1^2=1.59\), \(\sigma_2^2=1.83\), and \(\sigma_{12}=-0.22\), with random effects generated from the bivariate normal distribution, all five methods produced very similar RMSE values for both variance components when \((Se, Sp)=(0.50, 0.60)\) and \((0.70, 0.80)\). The RMSE decreased as the number of studies increased, and the differences among the competing methods became negligible for moderate and large meta-analyses.

More noticeable differences among the competing methods were observed in the high diagnostic accuracy setting \((Se, Sp)=(0.95, 0.99)\), particularly when only five studies were available. In this setting, the PB approach exhibited substantially larger RMSE values for both \(\sigma_1^2\) and \(\sigma_2^2\). The BGLMM also showed relatively higher RMSE for \(\sigma_2^2\), whereas the BLMM, NPB, and BCPB approaches produced comparatively lower RMSE values. As the number of studies increased, the RMSE values of all competing methods decreased rapidly, and the differences among methods became negligible.

\begin{figure}[ht]
    \centering
    \includegraphics[width=\linewidth]{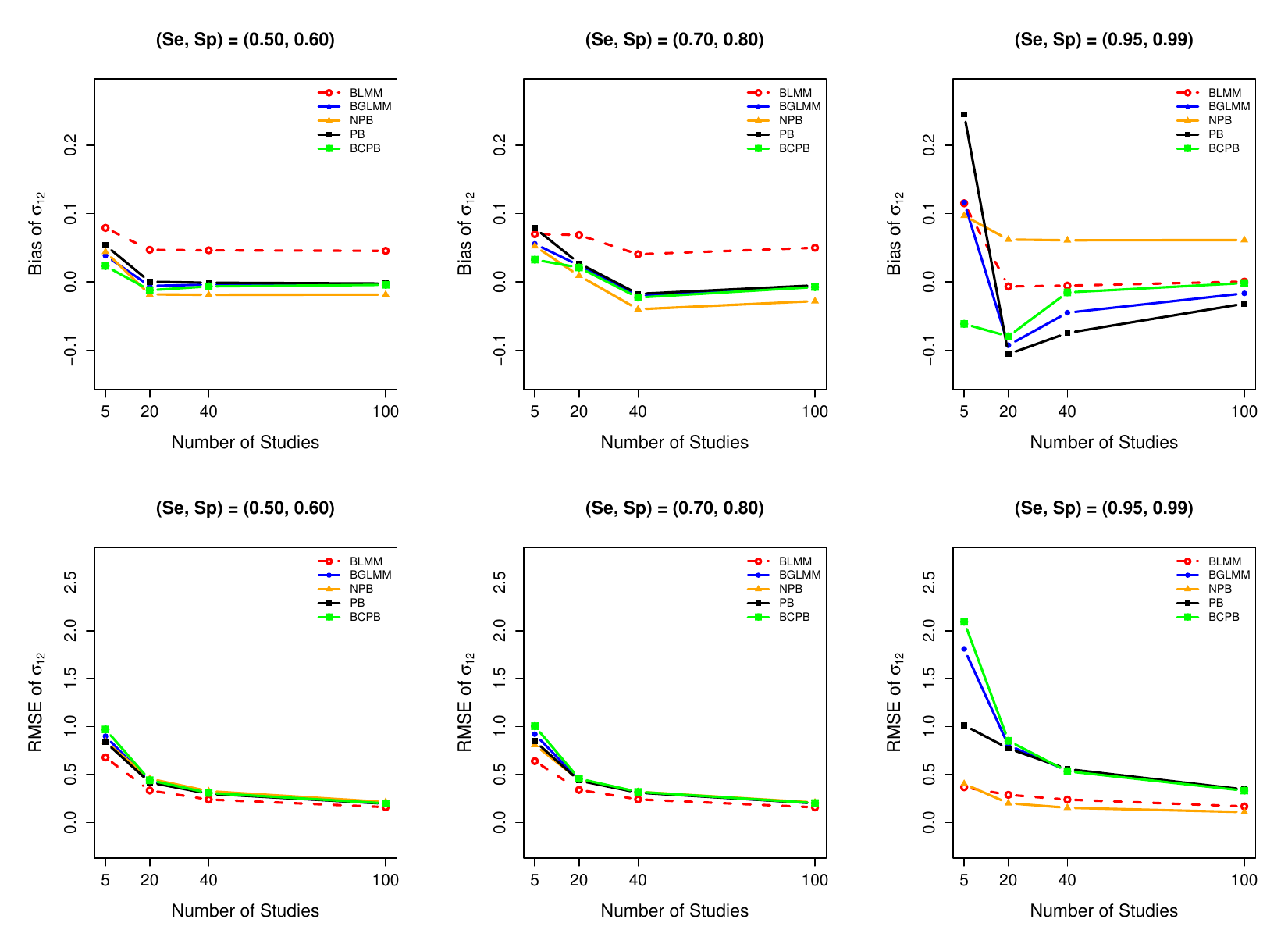}
    \caption{ Bias and RMSE of $\sigma_{12}$ for different combinations of studies and true ($Se$, $Sp$) pairs when $\sigma_1^2=1.59$, $\sigma_{12}=-0.22$, 
    $\sigma_2^2=1.83$, $n_1=20$, $n_2=50$ and random effects were generated from the normal distribution.}
    \label{fig_4.7}
\end{figure}

Figure~\ref{fig_4.7} presents the bias and RMSE of the between-study covariance estimate, \(\sigma_{12}\), under the setting \(n_1=20\), \(n_2=50\), \(\sigma_1^2=1.59\), \(\sigma_2^2=1.83\), and \(\sigma_{12}=-0.22\), with random effects generated from the bivariate normal distribution. For the \((Se, Sp)=(0.50, 0.60)\) and \((0.70, 0.80)\) settings, all five methods produced essentially unbiased estimates of $\sigma_{12}$ across all study sizes.

Under the high diagnostic accuracy setting \((Se, Sp)=(0.95, 0.99)\), the methods generally produced similar bias when estimating $\sigma_{12}$ for moderate and large numbers of studies. The main difference was observed when $k=5$, where the PB showed higher positive bias, whereas the proposed BCPB approach produced an estimate with bias closest to zero. As the number of studies increased, the bias values of the competing methods became very similar and remained close to zero.

For the RMSE of \(\sigma_{12}\), all five methods exhibited similar performance under the \((Se, Sp)=(0.50, 0.60)\) and \((0.70, 0.80)\) settings, with RMSE decreasing as the number of studies increased. In the high diagnostic accuracy setting, the proposed BCPB approach exhibited the largest RMSE when only five studies were available. However, its RMSE decreased sharply as the number of studies increased and became comparable to, or lower than, those of the competing methods. The BGLMM and PB approaches also exhibited elevated RMSE at $k=5$, although to a lesser extent than the BCPB, whereas the BLMM and NPB showed more stable RMSE across study sizes.

Following Zhao \textit{et al.} (2024),\cite{zhao2024evaluation} we assessed the convergence rate of each method, defined as the proportion of successfully converged model fits among the total number of fitted models in the simulation study. As shown in Table~\ref{A.3}, all five methods demonstrated excellent numerical stability across the bivariate Student-t, bivariate Normal, and bivariate Laplace random-effects distributions. The BLMM and NPB approaches achieved 100\% convergence across all three distributions. The BGLMM, PB, and BCPB approaches also exhibited excellent convergence, with convergence rates exceeding 98\% across all simulation settings. As expected, the PB and BCPB approaches shared the same convergence rates as the BGLMM because both methods rely on an initial successful BGLMM fit before bootstrap resampling. Overall, these results indicate that convergence was not a major concern for any of the competing methods under the simulation settings considered, because BGLMM was fitted using AGHQ, while models that failed to converge were sequentially refitted using the LA and IRLS algorithms.

\begin{table}[ht]
\centering
\caption{Convergence rates (\%) across all simulation scenarios under different random-effects distributions.}
\label{A.3}
\begin{tabular}{@{}lccc@{}}
\toprule
\textbf{Model} & \textbf{Student $t$ distribution} & \textbf{Bivariate Normal distribution} & \textbf{Bivariate Laplace distribution} \\
\midrule
BLMM   & 100.000\% & 100.000\% & 100.000\% \\
BGLMM  &99.318\% & 99.047\%  & 98.485\%  \\
NPB    & 100.000\% & 100.000\% & 100.000\% \\
PB     & 99.318\% & 99.047\% & 98.485\% \\
BCPB     & 99.318\% & 99.047\% & 98.485\% \\
\bottomrule
\end{tabular}
\end{table}

Overall, the proposed BCPB approach demonstrated strong finite-sample performance across most simulation settings, particularly for estimating sensitivity and specificity, while maintaining empirical CPs close to the nominal 95\% level. The BGLMM and PB approaches also exhibited excellent statistical performance, particularly for moderate- and large-sized meta-analyses. The proposed NPB approach generally improved upon the BLMM and provided a robust alternative without requiring parametric assumptions for the random-effects distribution. Similar conclusions were observed when estimating the between-study heterogeneity parameters, although differences among the competing methods became more apparent in the high diagnostic accuracy setting for very small meta-analyses. In general, the patterns for bias, RMSE, CI width, and empirical CP were broadly consistent across the three random-effects distributions, indicating that the relative performance of the competing methods was robust to deviations from normality.

\section{Illustrative Examples}
\label{s_6}

This section illustrates the methods discussed in this paper by using two real-life datasets briefly presented in Section \ref{s_2}. The datasets were chosen to demonstrate scenarios in which standard models may perform poorly (i.e., small meta-analyses and studies with skewed or unknown distributions). It is essential to note that, in order to estimate the pooled $Se$ and $Sp$ for both datasets using the BLMM and NPB models, the Haldane-Anscombe correction\cite{haldane1956estimation,anscombe1956estimating} was consistently applied whenever a cell contained zero counts, ensuring that the estimates and their variances remained valid and that the models converged.

\subsection{Results of the TEE Data}

Table \ref{t_4.1} presents the pooled sensitivity ($Se$), specificity ($Sp$), and between-study variance-covariance estimates obtained from the BLMM, BGLMM, NPB, PB, and BCPB models for the TEE dataset. All five methods indicate that the diagnostic test has low sensitivity but very high specificity, suggesting that it is considerably more effective at correctly identifying non-diseased participants than diseased ones.

\begin{table}[ht]
\centering
\caption{Comparison of the estimates of $Se$, $Sp$, and between-study heterogeneity parameters obtained from the BLMM, BGLMM, NPB, PB, and BCPB methods using the TEE dataset.}
\label{t_4.1}

\begin{tabular}{lccccc}
\toprule
\textbf{Model} & \textbf{Se (95\% CI)} & \textbf{Sp (95\% CI)} & $\sigma_1^2$ & $\sigma_2^2$ & $\sigma_{12}$ \\
\midrule
BLMM               & 21.78 (14.56, 31.26) & 98.53 (96.15, 99.45)  & 0.000 & 0.000 & 0.000 \\
BGLMM        & 18.95 (11.63, 29.33) & 99.53 (96.56, 99.94)  & 0.101 & 0.583 & -0.240 \\
NPB                & 20.59 (14.02, 27.59) & 98.49 (97.57, 99.11)  & 0.447 & 0.553 & 0.170 \\
PB           & 18.75 (11.18, 28.15) & 99.94 (98.46, 100.00) & 0.190 & 1.334 & -0.221 \\
BCPB        & 18.69 (9.64, 27.75) & 99.59 (98.71, 100.00) & 0.265 & 0.254 & -0.260 \\
\bottomrule
\end{tabular}
\begin{flushleft}
\footnotesize
\end{flushleft}

\end{table}

The estimated sensitivity ranges from 18.69\% under the BCPB model to 21.78\% under the BLMM, whereas the estimated specificity ranges from 98.49\% under the NPB model to 99.94\% under the PB. The BGLMM failed to converge when fitted using the LA and IRLS algorithms because the between-study variance--covariance structure could not be estimated reliably, preventing the likelihood optimization from reaching a stable solution. These numerical difficulties are likely attributable to the small number of studies and the sparse nature of the data, which included several zero cell counts. Specifically, the TEE dataset comprised only six studies, of which four had zero true positives, and one had zero false positives. Such sparse data can lead to instability in estimating the random-effects variance--covariance parameters, thereby hindering convergence of the optimization algorithm when using LA or IRLS. In contrast, the model converged successfully when fitted using the AGHQ algorithm, which provides a more accurate approximation to the marginal likelihood and is generally more robust in the presence of sparse data and complex random-effects structures.

The BLMM produced the highest sensitivity estimate (21.78\%) along with a similarly high specificity estimate (98.53\%), but all estimated variance components ($\sigma_1^2$, $\sigma_2^2$, and $\sigma_{12}$) were zero, which suggests that the model does not detect any between-study heterogeneity. In contrast, the BGLMM, NPB, PB, and BCPB methods yielded nonzero variance estimates, indicating between-study variability. Among these methods, the PB approach produced the highest estimate of specificity heterogeneity, while the standard BGLMM and NPB methods also identified substantial heterogeneity.

The estimated covariance between sensitivity and specificity differs across models. The BGLMM, PB and BCPB methods yielded negative covariance estimates, suggesting a potential trade-off between sensitivity and specificity across studies. In contrast, the NPB method produced a positive covariance estimate. This positive association likely reflects the small number of studies and the empirical dependence structure of the TEE dataset, in which studies with higher sensitivity also tend to exhibit higher specificity. Because the NPB resamples entire studies, these study-level patterns are preserved and may be amplified, resulting in an estimated positive covariance. Finally, the CIs for sensitivity were noticeably wider than those for specificity across all methods, indicating greater uncertainty in estimating sensitivity than specificity for the TEE dataset.

\subsection{Results of the EUS Data}

Table \ref{t_4.2} presents the pooled sensitivity and specificity estimates, together with corresponding CIs, and between-study variance components for all methods applied to the EUS dataset.

\begin{table}[ht]
    \centering
    \caption{Comparison of the estimates of $Se$, $Sp$, and between-study heterogeneity parameters obtained from the BLMM, BGLMM, NPB, PB, and BCPB methods using the EUS dataset.} 
     \label{t_4.2}
\vspace{10pt} % Add vertical space between the caption and the table\\
    \begin{tabular}{lccccc}
        \toprule
       \textbf{Model} & \textbf{Se (95\% CI)} & \textbf{Sp (95\% CI)} & \textbf{$\sigma_1^2 $} & \textbf{$\sigma_2^2$} & \textbf{$\sigma_{12}$} \\
        \midrule
    BLMM & 80.98 (76.90, 84.48) &  66.91 (61.87, 71.58)   &   0.380&  0.386& -0.262 \\
   BGLMM  & 83.21 (79.12, 86.63) &  66.58 (60.72, 71.96)  &  0.484 & 0.570 &-0.219 \\
   NPB  & 82.26 (78.67, 85.53) &  66.43 (60.72, 71.10)  &  0.922  &1.153 &-0.086  \\  
    PB  & 83.30 (79.45, 86.89) &  66.42 (60.34, 71.31)  &   0.479 &  0.553& -0.229 \\
   BCPB & 83.22 (79.26, 87.17) &   66.80 (61.08, 72.51) &  0.489 & 0.587 &-0.209  \\
        \bottomrule
    \end{tabular}
\end{table} 

The five methods yielded highly consistent pooled estimates for both sensitivity and specificity. The estimated sensitivity ranged from 80.98\% (BLMM) to 83.30\% (PB), while specificity varied minimally, from 66.42\% (PB) to 66.91\% (BLMM). These small differences indicate that the choice of model had limited impact on the point estimates of diagnostic accuracy for this dataset, suggesting that the overall conclusions regarding test performance are robust when the number of studies is large and the data are not sparse.

The estimates of between-study heterogeneity further highlighted structural differences among the models. The NPB method produced substantially higher variance components for both sensitivity and specificity compared with the other models. In addition, the estimated covariance between sensitivity and specificity under NPB was considerably closer to zero, indicating a weaker negative association between the two outcomes. By contrast, the BLMM, BGLMM, PB, and BCPB models all preserve a moderate negative dependence structure, consistent with the expected trade-off between sensitivity and specificity in diagnostic accuracy studies.

\subsection{Computing Time}
Table \ref{t_4.3}  summarizes the average computational time required by the five competing methods for analyzing the TEE and EUS data sets. The computing times were reported as user CPU time, system CPU time, and total elapsed time in the \texttt{R} software environment.\cite{R2025}

\begin{table}[ht]
    \centering
    \caption{ Average computing time (in seconds) of the five methods based on the two motivating data sets.}  
     \label{t_4.3}
    \begin{tabular}{lcccc}
        \toprule
     \textbf{Data} & \textbf{Model} & \textbf{User} & \textbf{System} & \textbf{Elapsed} \\
        \midrule
TEE data&   BLMM &  0.02 &  0.00  & 0.03 \\
        &  BGLMM  & 0.35 & 0.01  &  0.36 \\
      &  NPB  & 0.16 & 0.00  & 0.16 \\
      &  PB & 70.55 &  2.78 & 76.53  \\
       & BCPB & 75.26  & 3.04 & 78.98 \\
      \midrule
EUS data &  BLMM  &  0.04 &  0.00  & 0.04  \\
        &  BGLMM  & 0.21   & 0.00  &  0.25  \\
       &  NPB & 0.39 & 0.00  & 0.47  \\
      &   PB  & 39.94 & 1.22  & 46.39 \\
       & BCPB & 39.44  & 1.11 & 40.81 \\
        \bottomrule
    \end{tabular}
\end{table}

The results suggest that the BLMM, BGLMM, and NPB methods were computationally efficient, completing model fitting in a few seconds for both data sets, whereas the PB and BCPB methods required longer computation times than the other approaches due to the bootstrap resampling procedure.

\section{Discussion and Conclusion}
\label{s_7}
In this paper, we proposed the NPB, PB, and BCPB approaches for meta-analysis of DTA studies and compared their performance with the standard BLMM and BGLMM using extensive simulation studies and two real-life data applications. The proposed methods were motivated by practical challenges frequently encountered in DTA meta-analysis, including small numbers of studies, sparse data, and uncertainty regarding the underlying random-effects distribution.

Overall, the simulation results demonstrated that the BGLMM and the proposed bootstrap approaches achieved excellent statistical performance across a wide range of simulation settings. Consistent with previous studies, the BGLMM produced low bias, low RMSE, and empirical CPs close to the nominal level for pooled sensitivity and specificity estimates, reflecting the advantages of modelling the observed counts directly through binomial distributions rather than relying on normal approximations, particularly when diagnostic accuracy is high or sparse data are present.\cite{chu_bivariate_2006,zhao2024evaluation,rosenberger2022empirical} Building upon the BGLMM, the proposed BCPB approach further improved finite-sample estimation by reducing bias and RMSE for the pooled sensitivity and specificity estimates while maintaining empirical coverage probabilities close to the nominal level. These improvements were most evident in high diagnostic accuracy settings and in small meta-analyses, where finite-sample bias was more pronounced.

The proposed bootstrap approaches offered several practical advantages over the standard BLMM. Both the PB and BCPB approaches substantially reduced bias and RMSE for the pooled sensitivity and specificity estimates while maintaining empirical CPs close to the nominal level. In particular, the BCPB approach further improved upon the PB through bootstrap bias correction, resulting in better finite-sample performance, especially under high diagnostic accuracy settings and in small meta-analyses. The PB approach also demonstrated competitive performance compared to the standard BLMM. In contrast, the NPB approach provides a flexible alternative that avoids parametric assumptions for the random-effects distribution while maintaining competitive statistical performance across a broad range of simulation settings.

The improved finite-sample performance of the proposed bootstrap approaches is largely attributable to their ability to more closely approximate the finite-sample sampling distribution of the pooled sensitivity and specificity estimators. By generating bootstrap samples from the fitted BGLMM, the PB approach preserves the hierarchical structure of the observed data while reducing the reliance on asymptotic approximations. The additional bias-correction step incorporated into the proposed BCPB approach further reduced finite-sample bias, leading to improved estimation accuracy, particularly in high diagnostic accuracy settings and in small meta-analyses where finite-sample effects were most pronounced. These findings suggest that bootstrap-based estimation can provide a practical means of improving inference without altering the underlying bivariate modelling framework.

The simulation results also showed that estimation of the between-study heterogeneity parameters became considerably more challenging when the true sensitivity and specificity were very high and the number of studies was small. Under these settings, all competing methods experienced some loss of accuracy, particularly for the variance components. Nevertheless, the BGLMM, PB, and BCPB approaches generally provided more accurate estimates than the BLMM and NPB, with differences among the competing methods diminishing as the number of studies increased. For the between-study covariance, all methods produced comparable estimates for moderate and large meta-analyses, with noticeable differences observed only when the number of studies was very small.

The analyses of the TEE and EUS datasets were largely consistent with the simulation findings. All five methods produced comparable pooled sensitivity and specificity estimates for the EUS dataset, indicating that differences among methods become relatively small when the number of studies is sufficiently large. For the TEE dataset, which consisted of only six studies, the NPB approach produced stable estimates and converged successfully, illustrating its practical usefulness in small meta-analyses where conventional likelihood-based estimation may become computationally challenging.

Although the proposed bootstrap approaches demonstrated encouraging finite-sample performance, several limitations should be acknowledged. First, only three random-effects distributions were considered in the simulation study. Although these distributions represent a range of symmetric and heavy-tailed settings, future investigations involving more flexible distributions, such as skew-normal, skew-\(t\), or mixture distributions, may provide additional insight into the robustness of the proposed bootstrap approaches when the assumption of normal random effects is violated.\cite{negeri_skew-normal_2020,negeri2025bivariate} Second, the simulation study focused on summary-level DTA data and did not consider individual participant data meta-analysis (IPDMA) settings. Third, CIs for the proposed BCPB approach were constructed using bias-corrected Wald bootstrap procedures. Alternative bootstrap CI methods, such as bias-corrected and accelerated (BCa) intervals, may further improve finite-sample inference and warrant future investigation.\cite{efron_5_1982,davison1997bootstrap} Future work could also evaluate the performance of the proposed methods in the presence of study-level covariates, missing data, and other complexities commonly encountered in applied DTA meta-analyses.

Overall, this study demonstrates that bootstrap-based estimation procedures provide effective enhancements to existing likelihood-based methods for the meta-analysis of DTA studies. Across a wide range of simulation settings, the proposed bootstrap approaches demonstrated robust finite-sample performance and were resilient to deviations from the normality assumption for the random effects. Among the proposed methods, the BCPB approach consistently delivered the best overall performance for estimating pooled sensitivity and specificity by reducing finite-sample bias while maintaining low RMSE and empirical CPs close to the nominal level, particularly in high diagnostic accuracy settings and in small meta-analyses. The PB approach also exhibited strong statistical performance and produced results comparable to those of the BGLMM across most scenarios. In contrast, the NPB approach offers a flexible alternative that avoids parametric assumptions for the random-effects distribution while maintaining competitive statistical performance. Therefore, the proposed BCPB approach is a competitive approach for routine meta-analysis of DTAs due to its improved finite-sample performance. The NPB approach may be particularly useful when the underlying random-effects distribution is uncertain or when a distribution-free estimation procedure is preferred.

\paragraph{Acknowledgments.}
The authors are grateful for the computing facilities provided by the Faculty of Mathematics at the University of Waterloo.

\paragraph{Funding Statement.}
ZFN would like to acknowledge funding from the Natural Sciences and Engineering Research Council of Canada (NSERC), grant number RGPIN-2024-04652.

\paragraph{Author Contributions}
D.T.B. and O.A.D.: Investigation, methodology, software, formal analysis, visualization, writing—original draft. I.F.: Investigation, methodology and writing—original draft.  Z.F.N.: Conceptualization, investigation, methodology, software, writing—review and editing, funding acquisition, supervision. All authors approved the final submitted draft.

\bibliographystyle{unsrtnat}
\bibliography{references}  %%% Uncomment this line and comment out the ``thebibliography'' section below to use the external .bib file (using bibtex) .

@book{borenstein2021introduction,
  author = {Borenstein, M. and Hedges, V. L. and Higgins, P. T. J. and Rothstein, R. H.},
  title = {Introduction to Meta-Analysis},
  publisher = {John Wiley \& Sons},
  year = {2021},
  doi = {10.1002/9780470743386}
}

@phdthesis{negeri2019contributions,
  author = {Negeri, Z. F.},
  title = {Contributions to Statistical Methods for Meta-Analysis of Diagnostic Test Accuracy Studies},
  year = {2019},
  url = {https://macsphere.mcmaster.ca/items/d4fa7025-89f9-48f8-bf13-1256731d9aab},
 urldate = {2025-02-01}
}

@article{reitsma2005bivariate,
  author = {Reitsma, J. B.  and Glas, A. S. and Rutjes,  A. W. S. 
            and Scholten, R.J. P. M.  and Bossuyt, P. M. and Zwinderman, A. H.},
  title = {Bivariate Analysis of Sensitivity and Specificity Produces Informative Summary Measures in Diagnostic Reviews},
  journal = {Journal of Clinical Epidemiology},
  publisher = {Elsevier},
  year = {2005},
  volume = {58},
  number = {10},
  pages = {982--990},
  url = {https://doi-org.proxy.lib.uwaterloo.ca/10.1016/j.jclinepi.2005.02.022},
  urldate = {2025-01-01}
}

@article{borenstein2009effect,
  author = {Borenstein, M. and Cooper, H and Hedges, L. and Valentine, J.},
  title = {Effect Sizes for Continuous Data},
  journal = {The Handbook of Research Synthesis and Meta-analysis},
  year = {2009},
  volume = {2},
  pages = {221--235},
  url = {https://doi.org/10.7758/9781610448864},
  urldate = {2025-04-01}
}

@article{negeri_skew-normal_2020,
  author = {Negeri, Z. F. and Beyene, J.},
  title = {Skew-Normal Random-Effects Model for Meta-Analysis of Diagnostic Test Accuracy ({DTA}) Studies},
  journal = {Biometrical Journal},
  year = {2020},
  volume = {62},
  number = {5},
  pages = {1223--1244},
  doi = {10.1002/bimj.201900184}
}

@article{negeri_bivariate_2018,
  author = {Negeri, Z. F. and Mateen, S. and Beyene, J.},
  title = {Bivariate Random-Effects Meta-Analysis Models for Diagnostic Test Accuracy Studies Using Arcsine-Based Transformations},
  journal = {Biometrical Journal},
  year = {2018},
  volume = {60},
  number = {4},
  pages = {827--844},
  doi = {10.1002/bimj.201700101}
}

@article{negeri2025bivariate,
  author={Negeri, Zelalem F},
  title={A Bivariate Finite Mixture Random Effects Model for Identifying and Accommodating Outliers in Diagnostic Test Accuracy Meta-Analyses},
  journal={Biometrical Journal},
  year={2025},
  volume={67},
  number={3},
  pages={e70062},
 doi = {10.1002/bimj.70062}
}

@article{chu_bivariate_2006,
  author = {Chu, H. and Stephen, C. R.},
  title = {Bivariate Meta-Analysis of Sensitivity and Specificity with Sparse Data: A Generalized Linear Mixed Model Approach},
  journal = {Journal of Clinical Epidemiology},
  year = {2006},
  volume = {59},
  number = {12},
  pages = {1331--1332},
  doi = {10.1016/j.jclinepi.2006.06.011}
}

@article{noortgate_parametric_2005,
  author = {Noortgate, D. W. V. and Patrick, O.},
  title = {Parametric and Nonparametric Bootstrap Methods for Meta-Analysis},
  journal = {Behavior Research Methods},
  year = {2005},
  volume = {37},
  number = {1},
  pages = {11--22},
  url = {https://ocul-wtl.primo.exlibrisgroup.com/view/action/uresolver.do?operation=resolveService&package_service_id=13929874440005162&institutionId=5162&customerId=5150&VE=true},
  urldate = {2025-03-01}
}

@article{van_zaane_meta-analysis_2008,
  author = {Zaane, V. B. and Zuithoff, N. P. A.  and Reitsma, J. B. 
            and Bax, L. and Nierich, A. P. and Moons, K. G. M. },
  title = {Meta-Analysis of the Diagnostic Accuracy of Transesophageal Echocardiography for Assessment of Atherosclerosis in the Ascending Aorta in Patients Undergoing Cardiac Surgery},
  journal = {Acta Anaesthesiologica Scandinavica},
  year = {2008},
  volume = {52},
  number = {9},
  pages = {1179--1187},
  doi = {10.1111/j.1399-6576.2008.01694.x}
}

@article{efron_leisurely_1983,
  author = {Bradley, E. and Gail, G.},
  title = {A {Leisurely} {Look} at the {Bootstrap}, the {Jackknife}, and {Cross}-{Validation}},
  journal = {The American Statistician},
  year = {1983},
  volume = {37},
  number = {1},
  pages = {36--48},
  doi = {10.2307/2685844}
}

@article{zapf_nonparametric_2015,
  author = {Zapf, A. and Annika, H. and Katharina, K. and Oliver, K.},
  title = {Nonparametric Meta-Analysis for Diagnostic Accuracy Studies},
  journal = {Statistics in Medicine},
  year = {2015},
  volume = {34},
  number = {29},
  pages = {3831--3841},
  doi = {10.1002/sim.6583}
}

@article{burton_design_2006,
  author = {Burton, A. and Douglas, A. G. and Royston, P. and Holder, R. L.},
  title = {The Design of Simulation Studies in Medical Statistics},
  journal = {Statistics in Medicine},
  year = {2006},
  volume = {25},
  number = {24},
  pages = {4279--4292},
  doi = {10.1002/sim.2673}
}

@article{morris_using_2019,
  author = {Morris, T. P. and Ian, W. R. and Crowther, M. J. },
  title = {Using Simulation Studies to Evaluate Statistical Methods},
  journal = {Statistics in Medicine},
  year = {2019},
  volume = {38},
  number = {11},
  pages = {2074--2102},
  doi = {10.1002/sim.8086}
}

@article{mocellin_diagnostic_nodate,
  author = {Mocellin, S. and Pasquali, S.},
  title = {Diagnostic Accuracy of Endoscopic Ultrasonography ({EUS}) for the Preoperative Locoregional Staging of Primary Gastric Cancer - {Mocellin}, {S} - 2015 | {Cochrane} {Library}},
  journal = {Cochrane Library},
  issn = {1465-1858},
 url ={https://www.cochranelibrary.com/cdsr/doi/10.1002/14651858.CD009944.pub2/full},
 urldate = {2025-01-06}
}

@article{glass_primary_1976,
  author = {Glass, G. V. },
  title = {Primary, {Secondary}, and {Meta}-{Analysis} of {Research}},
  journal = {Educational Researcher},
  year = {1976},
  volume = {5},
  number = {10},
  pages = {3--8},
  doi = {10.3102/0013189X005010003}
}

@article{raudenbush_empirical_1985,
  author = {Raudenbush, S. W. and Bryk, A. S.},
  title = {Empirical {Bayes} {Meta}-{Analysis}},
  journal = {Journal of Educational Statistics},
  year = {1985},
  volume = {10},
  number = {2},
  pages = {75--98},
  doi = {10.3102/10769986010002075}
}

@article{negeri_statistical_2020,
  author = {Negeri, Z. F. and Beyene, J.},
  title = {Statistical Methods for Detecting Outlying and Influential Studies in Meta-Analysis of Diagnostic Test Accuracy Studies},
  journal = {Statistical Methods in Medical Research},
  year = {2020},
  volume = {29},
  number = {4},
  pages = {1227--1242},
  doi = {10.1177/0962280219852747}
}

@article{negeri_robust_2020,
  author = {Negeri, Z. F. and Beyene, J.},
  title = {Robust Bivariate Random-Effects Model for Accommodating Outlying and Influential Studies in Meta-Analysis of Diagnostic Test Accuracy Studies},
  journal = {Statistical Methods in Medical Research},
  year = {2020},
  volume = {29},
  number = {11},
  pages = {3308--3325},
  doi = {10.1177/0962280220925840}
}

@misc{doebler_mada_2025,
  author = {Doebler, P. and Sousa-Pinto, B.},
  title = {mada: {Meta}-{Analysis} of {Diagnostic} {Accuracy}},
  year = {2025},
  url = {https://cran.r-project.org/web/packages/mada/index.html},
 urldate = {2025-01-01}
}

@book{bryk_hierarchical_nodate,
  author = {Bryk, A. S.},
  title = {Hierarchical Linear Models: Applications and Data Analysis Methods},
  publisher = {Sage},
  address = {Newbury Park, Calif},
  series = {Advanced quantitative techniques in the social sciences},
  number = {1},
  isbn = {978-0-8039-4627-9},
  url ={https://books.google.ca/books?id=uyCV0CNGDLQC&printsec=frontcover&source=gbs_ge_summary_r&cad=0#v=onepage&q&f=false},
  urldate = {2025-05-12}
}

@incollection{goldstein_bootstrapping_2011,
  author = {Goldstein, H.},
  title = {Bootstrapping in Multilevel Models},
  booktitle = {Handbook for Advanced Multilevel Analysis},
  publisher = {Routledge/Taylor \& Francis Group},
  address = {New York, NY, US},
  year = {2011},
  pages = {163--171}, 
  url = {https://doi.org/10.1111/j.1467-985X.2011.00709_6.x},
  urldate = {2025-02-12}
}

@article{laird_empirical_1987,
  author = {Laird, N. M. and Louis, T. A.},
  title = {Empirical {Bayes} {Confidence} {Intervals} {Based} on {Bootstrap} {Samples}},
  journal = {Journal of the American Statistical Association},
  year = {1987},
  volume = {82},
  number = {399},
  pages = {739--750},
  doi = {10.2307/2288778}
}

@incollection{efron_5_1982,
  author = {Bradley, E.},
  title = {5. {The} {Bootstrap}},
  booktitle = {The {Jackknife}, the {Bootstrap} and {Other} {Resampling} {Plans}},
  publisher = {Society for Industrial and Applied Mathematics},
  year = {1982},
  pages = {27--36},
  doi = {10.1137/1.9781611970319.ch5}
}

@article{viechtbauer2010conducting,
  author = {Viechtbauer, W.},
  title = {Conducting Meta-Analyses in {R} with the metafor Package},
  journal = {Journal of Statistical Software},
  year = {2010},
  volume = {36},
  pages = {1--48},
  doi = {10.18637/jss.v036.i03}
}

@article{veroniki2016methods,
  author = {Angeliki, V. A. and Dan, J. and Viechtbauer, W.
            and Bender, R. and Bowden, J. and Knapp, G. and Kuss, O.
            and Higgins, J. P. T. and Langan, D. and Salanti, G.},
  title = {Methods to Estimate the Between-Study Variance and Its Uncertainty in Meta-Analysis},
  journal = {Research Synthesis Methods},
  publisher = {Wiley Online Library},
  year = {2016},
  volume = {7},
  number = {1},
  pages = {55--79},
  doi = {10.1002/jrsm.116455}}

@article{haldane1956estimation,
  author    = {Haldane, J. B. S.},
  title     = {The estimation and significance of the logarithm of a ratio of frequencies},
  journal   = {Ann. Hum. Genet.},
  year      = {1956},
  volume    = {20},
  number    = {4},
  pages     = {309--311},
  doi       = {10.1111/j.1469-1809.1955.tb01285.x}
}

@article{anscombe1956estimating,
  author    = {Anscombe, F. J.},
  title     = {On Estimating Binomial Response Relations},
  journal   = {Biometrika},
  year      = {1956},
  volume    = {43},
  number    = {3/4},
  pages     = {461--464},
  doi       = {10.2307/2332926}
}

@article{bates2015fitting,
  author    = {Bates, D. and M{\"a}chler, M. and Bolker, B. and Walker, S.},
  title     = {Fitting Linear Mixed-Effects Models Using lme4},
  journal   = {J. Stat. Softw.},
  year      = {2015},
  volume    = {67},
  number    = {1},
  pages     = {1--48},
  doi       = {10.18637/jss.v067.i01}
}

@article{rosenberger2022empirical,
  author    = {Rosenberger, K. J. and Chu, H. and Lin, L.},
  title     = {Empirical comparisons of meta-analysis methods for diagnostic studies: a meta-epidemiological study},
  journal   = {BMJ Open},
  year      = {2022},
  volume    = {12},
  number    = {5},
  pages     = {e055336},
  doi       = {10.1136/bmjopen-2021-055336}
}

@article{dabi2026diagnostic,
  title={Diagnostic test accuracy meta-analysis based on an exact within-study variance calculation method},
  author={Dabi, Olana Angesa and Negeri, Zelalem Firisa},
  journal={Research Synthesis Methods},
  volume={17},
  number={4},
  pages={693--713},
  year={2026},
  publisher={Cambridge University Press}
}

@article{pinheiro1995approximations,
  author    = {Pinheiro, J. C. and Bates, D. M.},
  title     = {Approximations to the Log-Likelihood Function in the Nonlinear Mixed-Effects Model},
  journal   = {J. Comput. Graph. Stat.},
  year      = {1995},
  volume    = {4},
  number    = {1},
  pages     = {12--35},
  doi       = {10.2307/1390625}
}

@incollection{jorgensen2006iteratively,
  author    = {Jorgensen, Marit},
  title     = {Iteratively reweighted least squares},
  booktitle = {Encyclopedia of Environmetrics},
  year      = {2006},
  publisher = {Wiley},
  address   = {Chichester, UK},
  doi       = {10.1002/9780470057339}
}

@article{liu1994note,
  author    = {Liu, Q. and Pierce, D. A.},
  title     = {A Note on Gauss-Hermite Quadrature},
  journal   = {Biometrika},
  year      = {1994},
  volume    = {81},
  number    = {3},
  pages     = {624--629},
  doi       = {10.2307/2337136}
}

@article{zhao2024evaluation,
  author    = {Zhao, Y. and Khan, B. and Negeri, Z. F.},
  title     = {An evaluation of computational methods for aggregate data meta-analyses of diagnostic test accuracy studies},
  journal   = {BMC Med Res Methodol},
  year      = {2024},
  volume    = {24},
  number    = {1},
  pages     = {111},
  doi       = {10.1186/s12874-024-02217-2}
}

@Manual{R2025,
  title = {R: A Language and Environment for Statistical Computing},
  author = {{R Core Team}},
  organization = {R Foundation for Statistical Computing},
  address = {Vienna, Austria},
  year = {2025},
  url = {https://www.R-project.org/},
}

@Manual{statisticat2013laplacesdemon,
  title        = {LaplacesDemon: Complete Environment for Bayesian Inference},
  author       = {{Statisticat, LLC}},
  year         = {2013},
  note         = {R package version 16.1.8},
  url          = {https://github.com/LaplacesDemonR/LaplacesDemon}
}

@article{genz2021package,
  title={Package ‘mvtnorm’},
  author={Genz, Alan and Bretz, Frank and Miwa, Tetsuhisa and Mi, Xuefei and Leisch, Friedrich and Scheipl, Fabian and Bornkamp, Bjoern and Maechler, Martin and Hothorn, Torsten and Hothorn, Maintainer Torsten},
  journal={Journal of Computational and Graphical Statistics},
  volume={11},
  number={950-971},
  pages={155},
  year={2021}
}

@article{VanDenNoortgate2005,
  author  = {Van Den Noortgate, Wim and Onghena, Patrick},
  title   = {Parametric and Nonparametric Bootstrap Methods for Meta-Analysis},
  journal = {Behavior Research Methods},
  year    = {2005},
  volume  = {37},
  number  = {1},
  pages   = {11--22},
  doi     = {10.3758/BF03206394}
}

@article{tibshirani1993introduction,
author={Tibshirani, Robert J and Efron, Bradley},
  title={An introduction to the bootstrap},
  journal={Monographs on Statistics and Applied Probability},
  year={1993},
  volume={57},
  number={1},
  pages={1--436},
doi     = {10.1201/9780429246593}
}

@article{kuss2014meta,
  title={Meta-analysis for diagnostic accuracy studies: a new statistical model using beta-binomial distributions and bivariate copulas},
  author={Kuss, Oliver and Hoyer, Annika and Solms, Alexander},
  journal={Statistics in medicine},
  volume={33},
  number={1},
  pages={17--30},
  year={2014},
  publisher={Wiley Online Library}
}

@article{steck2003bias,
  title={Bias-corrected bootstrap and model uncertainty},
  author={Steck, Harald and Jaakkola, Tommi},
  journal={Advances in neural information processing systems},
  volume={16},
  year={2003}
}

@book{davison1997bootstrap,
  author    = {Davison, Anthony C. and Hinkley, David V.},
  title     = {Bootstrap Methods and Their Application},
  publisher = {Cambridge University Press},
  address   = {Cambridge},
  year      = {1997},
  isbn      = {9780521574716}
}

@Manual{R2026,
  title        = {R: A Language and Environment for Statistical Computing},
  author       = {{R Core Team}},
  organization = {R Foundation for Statistical Computing},
  address      = {Vienna, Austria},
  year         = {2026},
  url          = {https://www.R-project.org/},
}

@Manual{Rizopoulos2023,
  title  = {GLMMadaptive: Generalized Linear Mixed Models using Adaptive Gaussian Quadrature},
  author = {Dimitris Rizopoulos},
  year   = {2023},
  note   = {R package version 0.9-1},
  url    = {https://cran.r-project.org/package=GLMMadaptive}
}





% \end{thebibliography}

\end{document}